\documentclass[12pt]{article}

\usepackage[title]{appendix}
\usepackage{amsfonts}
\usepackage{amssymb}
\usepackage{amsthm}
\usepackage[fleqn]{amsmath}
\usepackage[mathcal]{euscript}
\usepackage{mathrsfs}

\usepackage[latin1,ansinew]{inputenc}
\usepackage{fullpage}
\usepackage{psfrag}
\usepackage{setspace}
\usepackage{multirow}
\usepackage{longtable}
\usepackage{graphicx}
\usepackage{booktabs}
\usepackage{rotating} 

\usepackage{float}

      \usepackage{array}
      \usepackage{calc}
      \usepackage{hhline}
      \usepackage{ifthen}

\input{epsf}
\graphicspath{{./Figures/}}

\usepackage[usenames,dvipsnames]{color}
\usepackage[usenames,dvipsnames]{colortbl}
\usepackage{hyperref}

\newcommand{\MARKED}[2]{{\textcolor{#1}{#2}}}

\DeclareFontFamily{U}{rcjhbltx}{}
\DeclareFontShape{U}{rcjhbltx}{m}{n}{<->rcjhbltx}{}
\DeclareSymbolFont{hebrewletters}{U}{rcjhbltx}{m}{n}

\let\aleph\relax\let\beth\relax
\let\gimel\relax\let\daleth\relax

\DeclareMathSymbol{\aleph}{\mathord}{hebrewletters}{39}
\DeclareMathSymbol{\beth}{\mathord}{hebrewletters}{98}
\DeclareMathSymbol{\gimel}{\mathord}{hebrewletters}{103}
\DeclareMathSymbol{\daleth}{\mathord}{hebrewletters}{100}

\DeclareMathSymbol{\lamed}{\mathord}{hebrewletters}{108}
\DeclareMathSymbol{\mem}{\mathord}{hebrewletters}{109}
\DeclareMathSymbol{\ayin}{\mathord}{hebrewletters}{96}
\DeclareMathSymbol{\tsadi}{\mathord}{hebrewletters}{118}
\DeclareMathSymbol{\qof}{\mathord}{hebrewletters}{114}
\DeclareMathSymbol{\shin}{\mathord}{hebrewletters}{152}

\newtheorem{corollary}{Corollary}
\newtheorem{definition}{Definition}
\newtheorem{lemma}{Lemma}
\newtheorem{proposition}{Proposition}
\newtheorem{remark}{Remark}
\newtheorem{theorem}{Theorem}

\newcommand{\Ref}[1]{\eqref{#1}}

\newcommand{\sign}{\mathrm{sign}\,}

\newcommand{\vect}[1] {\boldsymbol{{ #1}} }

\newcommand{\pV}{{\vect{p}}}           

\DeclareMathAlphabet{\mathpzc}{OT1}{pzc}{m}{it}

\newcommand{\dd}{\mathrm{d}}

\newcommand{\Eset}{\mathbb{E}}

\newcommand{\Nset}{\mathbb{N}}
\newcommand{\Rset}{\mathbb{R}}
\newcommand{\Sset}{\mathbb{S}}

\newcommand{\esp}{\mathfrak{e}}

\begin{document}

\title{$\!$Hilbert's `monkey saddle' and other curiosities in$\!$ \\
       the equilibrium problem of three point particles on a circle 
       for repulsive power law forces}
\author{\normalsize \sc{Michael K.-H. Kiessling and Renna Yi}\\[-0.1cm]
	\normalsize  Department of Mathematics, Rutgers University,\\[-0.1cm]
	\normalsize 110 Frelinghuysen Rd., Piscataway, NJ 08854, USA}
\vspace{-0.3cm}
\date{$\phantom{nix}$}
\maketitle
\vspace{-2cm}

\begin{abstract}
\noindent
        This article determines all possible (proper as well as pseudo) equilibrium arrangements 
under a repulsive power law force of three point particles on the unit circle.
 These are the critical points of the sum over the three (standardized) Riesz pair interaction terms,
each given by 
$V_s(r)= s^{-1}\left(r^{-s}-1 \right)$ when the real parameter $s \neq 0$, and by 
$V_0(r) := \lim_{s\to0}V_s(r) =  -\ln r$;
here, $r$ is the chordal distance between the particles in the pair.
        The bifurcation diagram which exhibits all these equilibrium arrangements together as functions of $s$
features three obvious ``universal'' equilibria, which do not depend on $s$, and two not-so-obvious continuous
families of $s$-dependent non-universal isosceles triangular equilibria. 
        The two continuous families of non-universal equilibria are disconnected, yet they bifurcate off of a common universal
limiting equilibrium (the equilateral triangular configuration), at $s=-4$, where the graph of the total Riesz energy 
of the 3-particle configurations has the shape of a ``monkey saddle.''
        In addition, one of the families of non-universal equilibria also bifurcates off of another universal 
equilibrium (the antipodal arrangement), at $s=-2$. 
        While the bifurcation at $s=-4$ is analytical, the one at $s=-2$ is not.
 The bifurcation analysis presented here is intended to serve as template for the treatment of similar $N$-point
equilibrium problems on $\Sset^d$ for small~$N$.
\end{abstract}

\vfill
\hrule
\smallskip\noindent
{\footnotesize
Typeset in \LaTeX\ by the authors.  Original: Dec. 14, 2017; revised: Dec.29, 2018.

\smallskip\noindent
\copyright 2018 The authors. This preprint may be reproduced for noncommercial purposes.}

	\section{Introduction}\vspace{-10pt}

\noindent
  The 7th entry in Stephen Smale's list of problems worthy of the attention of mathematicians in the 21st century
\cite{Smale} asks for an algorithm which in polynomial($N$) many steps returns an $N$-point configuration on 
the two-sphere $\Sset^2$, whose logarithmic energy does not deviate from the optimal value $\mathcal{E}(N)$ 
by more than the order of the fourth term in the partly rigorously established, partly only conjectured 
asymptotic large-$N$ expansion
\begin{equation}
\mathcal{E}(N)
=\label{asympCONJlogS2}
 a N^2 + b N\ln N +cN + d \ln N + \mathcal{O}(1),
\end{equation}
with $a= \textstyle{\frac{1}{4}}\ln\textstyle{\frac{e}{4}}$, $b = - \textstyle{\frac{1}{4}}$, and
$c = \ln\big(2(2/3)^{1/4}\pi^{3/4}/\Gamma(1/3)^{3/2}\big)$ (see \cite{BrHaSa2012,Betermin}), 
while the value of $d$ is still unsettled --- 
what matters in Smale's 7th problem is not the exact value of $d$ but that the discrepancy is $\mathcal{O}(\ln N)$.
 Subsequently Smale remarked that similar requests can be made with the logarithmic energy replaced by a
family of Riesz $s$-energies with $s\in(0,2)$ (the logarithmic energy ``morally'' standing for $s=0$), and 
that analogous problems can be stated with $\Sset^2$ replaced by $\Sset^d$, $d\in\Nset$.
   The Riesz $s$-energy of an $N$-point configuration on $\Sset^d$ is obtained by summing up the Riesz $s$-energies 
of the $N(N-1)/2$ pairs in an $N$-point configuration. 
   The Riesz $s$-energy $R_s(r)$ of a pair of points with position vectors $\pV_{1}\in\Rset^{d+1}$ and 
$\pV_{2}\in\Rset^{d+1}$ a Euclidean distance $\left|\pV_{1}-\pV_{2}\right|= r$ apart is 
defined as $R_s(r):=\sign(s) r^{-s}$ for $s\neq 0$ (in the convention of \cite{Sch2016arXiv})
and as $R_0(r):=-\ln r$ for $s=0$ (the ``logarithmic energy'').\footnote{The discontinuous jumps of
  $R_s(r)$ at $s=0$ cause some artificial difficulties when trying to compare optimal energies at negative, zero,
  and positive $s$ values. 
  This can be rectified by introducing an $s$-continuous ``standardized Riesz $s$-energy'' $V_s(r)$, see below.}
   An ``optimal energy configuration'' is an absolute minimizer of the configurational Riesz $s$-energy.
   For background reading concerning the quest for
optimal Riesz $s$-energy configurations of $N$ point particles on $\Sset^{2}$ and other manifolds, 
see the survey articles\footnote{Some of these references
  use the definition $R_s(r)=r^{-s}$ for $s\neq 0$, and $R_0(r)=-\ln r$ for $s=0$, e.g.
 \cite{SaffKuijlaars}.
   By an ``optimal energy configuration'' one then means a configuration which minimizes the configurational 
   Riesz $s$-energy when $s\geq 0$, or maximizes it when $s < 0$.
   The historical origin of ignoring the $\sign(s)$-factor in front of $r^{-s}$, which entails that one searches for 
   the ``maximum energy configuration'' when $s<0$ instead of the conventional ``minimum,'' seems to be that for 
   $s=-1$ the problem is identical to
   the maximal average pairwise distance problem, cf. \cite{FejToth,Stol,Beck}. 
   Yet, having to constantly distinguish between energy minimization for $s\geq 0$ and energy maximization for $s<0$ 
   in a physics-inspired narrative (i.e.: optimal ``energy'') is somewhat awkward.
    Incidentally, the physics origin of these optimal-energy problems seems to go back to a pre-quantum mechanics
    inquiry into the structure of atoms by J.J. Thomson \cite{Thomson}. 
    Thomson studied, among other things, the minimum energy configurations on the circle $\Sset^1$ of $N$ point electrons with 
    Coulomb's electric pair interactions (the case $s=1$ in $R_s(r)$), and he made a brief remark that similar
    questions can be asked for $N$ point electrons on the sphere $\Sset^2$.
    The $1/r$ problem on $\Sset^2$ was later dubbed ``Thomson problem'' by Whyte \cite{Whyte}, and the same problem with $1/r$
    replaced by $R_s(r)$ is sometimes called ``the generalized Thomson problem.''}
 \cite{ErberHockneyTWO}, \cite{SaffKuijlaars}, \cite{HardinSaffONE}, Appendix 1 in \cite{NBK}, 
the websites \cite{BCM} and \cite{Womersley}, and the related article \cite{AtiyahSutcliffe}.

 It is easy to convey a sense of the challenge posed by Smale's 7th problem, and by the quest for optimal Riesz $s$-energy 
$N$-point configurations in general, when $N$ is large. 
 The empirical number count of relative energy minimizers shows a roughly exponential increase with $N$, see
\cite{ErberHockneyTWO}. 
 In the absence of a deterministic algorithm which in polynomially($N$) many steps finds an
optimizing configuration, the method of choice has been (quasi-)random searches (with variations on this theme), 
yet the likelihood that such a search produces a relative minimizer which is not absolute increases dramatically with $N$.
 
 To rigorously prove, or disprove, the exponential increase with $N$ of the number of relative equilibria when $N$ is large is an 
interesting open problem. 
 As far as we know, it is not even rigorously known whether exponentially increasing upper or lower bounds 
on the number of relative equilibria exist.
 
 The overwhelmingly large number of relative minimizers which need to be avoided is not the sole
source of difficulties for reaching an optimizer.
 Even in the small $N$ regime, where it would seem reasonable to expect that it should not be difficult to identify all the relative 
minimizers of the Riesz $s$-energy function of $N$-particle arrangements\footnote{We will generally speak of 
  particle \emph{arrangements} on the unit circle because ``degenerate configurations''
     are not \emph{configurations} in the proper sense.}
 on $\Sset^d$, at least for small $d=1$ and $d=2$, 
and determine the optimizer(s) amongst them, it is not necessarily easy to work out the answers.
 To be sure, the $N=2$ particle problem on $\Sset^d$, $d\in\Nset$, is trivial:\footnote{And, of course, it is meaningless to ask for
   a minimum Riesz \emph{pair energy} configuration when $N=1$, since one cannot form a pair with a single particle alone.}
since the Riesz $s$-energy of a pair of particles decreases monotonically
with the distance $r$ between the particles in the pair, the energy of the pair is
minimized when the two particles are at their furthest possible distance from each other, which is the antipodal configuration. 
          It is clear that this is also the only relative minimizer (up to rotation) when $N=2$. 
 However, already the optimal $N=3$ Riesz $s$-energy arrangement on $\Sset^d$ is an interesting problem.

      Although the absolute Riesz $s$-energy minimizers with three particles have
been correctly stated (though without detailing the proof)\footnote{Incidentally, there is a slip of pen at the pertinent
   place in Appendix 1 of \cite{NBK}: ``the only equilibrium arrangements'' should read ``the only stable equilibrium arrangements.''\vspace{-20pt}}
in Appendix 1 of \cite{NBK}, to the best of our knowledge the complete list of critical Riesz $s$-energy arrangements of $N=3$ particles,
and their stability, has not previously been discussed rigorously.
      In this paper we will supply such a discussion.

 By a repulsive power-law force equilibrium on $\Sset^d$ we mean the following.

   The \emph{proper repulsive power-law force} of a point particle at $\pV'\in\Rset^{d+1}$ onto a point particle at
$\pV\in\Rset^{d+1}$ is defined as the negative $\pV$-gradient of the Riesz $s$-energy $R_s(\left|\pV-\pV'\right|)$,
whenever this gradient is well-defined; if not well-defined, we may still define a \emph{pseudo force}, see below.
   The total force on a point particle at $\pV\in\Rset^{d+1}$ is the sum of all these forces 
exerted on it by the other particles in an $N$-particle system.
   When restricted to move on $\Sset^d$, $d\in\Nset$, an arrangement of $N$ point particles is said to be
in \emph{(proper or pseudo) force equilibrium} on $\Sset^d$
if the $\pV$-gradient of the total Riesz $s$-energy points radially at each particle's position $\pV$.

 Such force equilibria with $N=3$ particles are the same on every $\Sset^d$ (after factoring out rotations). 
 Only their stability properties may depend on $d$.

 There are three $s$-independent equilibria, which we call \emph{universal}, and which
are obvious to everyone. 
 There are also two $s$-dependent continuous families of not-so-obvious equilibria, which we
call \emph{non-universal}.

  One universal repulsive force equilibrium with $N=3$ particles comes to mind immediately: the equilateral triangular
configuration is in proper force equilibrium for all $s\in\Rset$ due to its symmetry, and because it is
not degenerate.
   We will see that w.r.t. motions on $\Sset^1$, depending on $s$, the equilateral triangular 
configuration will be any of the following: 
relatively but not absolutely maximizing, a saddle point, only relatively or also absolutely minimizing.

   When $s<-1$ the Riesz $s$-energy gradient force of one point particle onto another one is well-defined also when both particles
occupy the same point in space, in which case the mutual forces vanishes.
   Thus, when $s<-1$, the completely degenerate configuration (i.e. all $N$ particles occupy the same point in space) is 
in proper force equilibrium, and so is any antipodal arrangement (likewise a degenerate configuration, but not completely)
with $1\leq n<N$ particles at the North, and $N-n$ particles at the South Pole (up to rotation). 
 When $N=3$, there is only one antipodal arrangement (up to rotation), say with $n=1$. 
 The completely degenerate configuration is manifestly the absolute energy maximizer, while the antipodal arrangements can
be relative or even absolute energy minimizers, or saddle points, depending on $s$. 

  When $s\geq -1$ the Riesz $s$-energy gradient is not well-defined when two particles occupy the same point, yet when $-1\leq s < 0$ 
the degenerate configurations are still absolutely Riesz $s$-energy maximizing, respectively saddles, and so we refer to
them as \emph{pseudo force equilibria} when $-1\leq s < 0$. 
  For $s\geq 0$ these degenerate configurations have $+\infty$ Riesz $s$-energy, so that the notion of them being absolute energy 
maxima or saddles becomes ambiguous; yet by defining the pseudo force of one particle on top of another as a Cauchy-type principal 
value, which vanishes, we will count the degenerate configurations (one-point, or antipodal) as non-proper pseudo
force equilibria when $s\geq 0$.

 The interesting part of the $N=3$ problem is the non-obvious existence of two continuous families of
non-universal isosceles repulsive power-law force equilibria whose shapes depend on the Riesz parameter $s$. 
 The isosceles families bifurcate off of a common endpoint --- the equilateral configuration --- at $s=-4$;
one of the families also bifurcates off of the antipodal arrangement, at $s=-2$. 
 The bifurcation at $s=-4$ is analytical, the one at $s=-2$ is not.
 The isosceles triangular configurations are saddle points w.r.t. motions on $\Sset^1$.
 
 We next state our results precisely, then prove them rigorously.\vspace{-20pt}
 
	\section{Statement of the Main Results}

\begin{figure}[H]
\centering
\includegraphics[scale=.5]{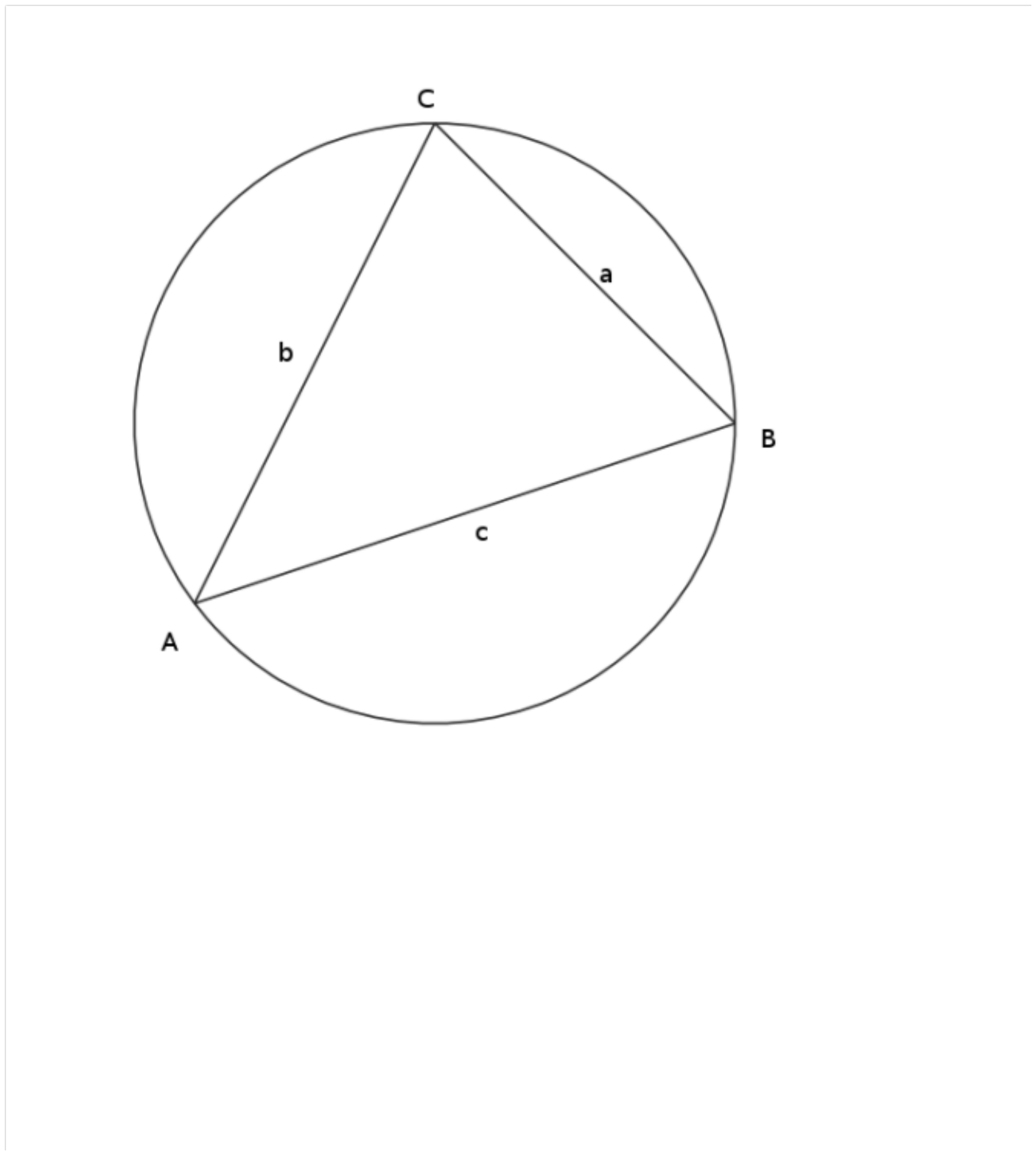} \vspace{-4truecm}
\caption{Notational conventions for a triangle and its circumcircle.}
\label{TinC}
\end{figure}

 Suppose $N=3$ point particles in Euclidean $\Eset^2$ are located on a circle of radius 1,  $\Sset^1\subset\Eset^2$,
forming the corners $\{A,B,C\}$ of a non-degenerate triangle with sides $\{a,b,c\}$ having lengths $\{|a|,|b|,|c|\}$, 
and strictly positive angles $\{\alpha,\beta,\gamma\}$, cf. Fig.\ref{TinC}.\vspace{-4truecm}
{\vskip-3.45truecm$\hspace{4.8truecm}\alpha$\vskip-1.6truecm$\hspace{8.2truecm}\beta$\vskip-2.2truecm$\hspace{6.2truecm}\gamma$}
\vspace{6truecm}

 An arrangement of three point particles on $\Sset^1$ labeled $(A,B,C)$ in counter-clockwise manner is characterized by a point 
$(\alpha,\beta,\gamma)\in [0,\pi]^3\cap \{\alpha+\beta+\gamma=\pi\}$, the ``fundamental triangle'' in $(\alpha,\beta,\gamma)$ space.
\begin{figure}[H]
\centering
\includegraphics[scale=.5]{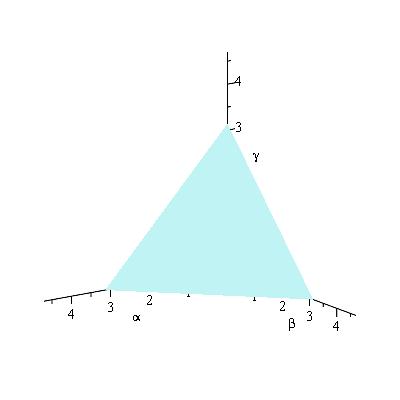} 
\caption{The fundamental triangle in $(\alpha,\beta,\gamma)$ space.}
\label{FT}
\end{figure}
 Since a triangular arrangement of point particles on $\Sset^1$ remains invariant under cyclic permutation of the counter-clockwise 
labelling of the points, each geometrical triangular arrangement of point particles is represented three times in the fundamental 
triangle, with the exception of the equilateral triangular configuration,
represented by the center $(\frac{\pi}{3},\frac{\pi}{3},\frac{\pi}{3})$ of the fundamental triangle. 

 All interior points of the fundamental triangle represent proper $3$-point configurations (no two point particles occupy the
same location). 
 Degenerate configurations with precisely two or all three point particles occupying the same location are represented by the
edges, respectively corners of the fundamental triangle. 
 Indeed, by inspecting Fig.\ref{TinC} is is clear that when precisely two point particles are being moved to the same location 
(say particle $B$ is moved to the location of particle $C$), then $\angle(C,A,B)=\alpha \to 0$. 
 But when $\alpha=0$, then $\beta+\gamma=\pi$ 
means the arrangement is represented by a point on the edge of the fundamental triangle which sits in the $(\beta,\gamma)$ plane. 
 Of special importance are the antipodal arrangements $(\alpha,\beta,\gamma)=(0,\frac{\pi}{2},\frac{\pi}{2})$ (up to permutation)
which are located at the midpoints of the edges of the fundamental triangle, and
the completely degenerate triangular configuration (i.e. the one-point arrangement) given by
$(\alpha,\beta,\gamma)= (0,0,\pi)$ (up to permutation).

 Let $r>0$ stand for any of the side lengths of a proper triangular configuration.
 We now assign each side a \emph{standardized Riesz $s$-energy} $V_{s}(r)$
\emph{of the pair of point particles} located at the two end points, defined by\footnote{We remark 
  (see Appendix 3 in \cite{NBK}) that $s\mapsto V_s(r)$ is a continuous, monotonically increasing function for all $r>0$,
  strictly so if $r\neq 1$, and $V_s(1)\equiv 0$ for all $s$.}
\begin{eqnarray}
	V_{s}(r) &:=&\label{RieszVs}
	s^{-1}\left(r^{-s}-1 \right), 	\qquad s\in \Rset, \quad s \neq 0;\\
	V_{0}(r) &:=& \label{RieszVnull}
	-\ln{r}                
\qquad\qquad\qquad \Big(\!\!=
\lim_{s\rightarrow 0} V_{s}(r)\Big),
\end{eqnarray}
and average these pair energies over the triangular configuration, written as
\begin{equation} 
	\langle V_s\rangle (\alpha,\beta,\gamma)
:=\label{aveRIESZpairENERGY}
	\tfrac13\big(V_s(|a|) + V_s(|b|) + V_s(|c|)\big);
\end{equation}	
here we tacitly used that $|a|=2\sin\alpha$, $|b|=2\sin\beta$, and $|c|=2\sin\gamma$.
 We define the \emph{minimal average standardized Riesz pair-energy} of $N=3$ particles on a unit circle as 
\begin{equation} 
	v_s^{}(3)
:=\label{MINaveRIESZpairENERGY}
	\inf_{\alpha+\beta+\gamma=\pi} \langle V_s\rangle(\alpha,\beta,\gamma).
\end{equation}		

 It is easy to see that the {minimal average standardized Riesz pair-energy} \Ref{MINaveRIESZpairENERGY}
exists for all $s\in\Rset$, but the infimum may not be achieved on the set of non-degenerate triangular configurations. 
 We now compactify the space of non-degenerate triangular configurations on the unit circle by adding to it
all degenerate three-particle configurations, viz. the linear arrangements in which precisely one of the three
angles $\alpha,\beta,\gamma$ is zero, and the single-point arrangement in which two of the three angles $\alpha,\beta,\gamma$ are 
zero.
 We also assign an extended real Riesz $s$-energy to a pair with vanishing distance $r=0$ in terms of 
$\lim_{r\downarrow 0} V_{s}(r)$, viz.
\begin{eqnarray}
	V_{s}(0) &:=&\label{RieszVsATnull}
	-s^{-1} ,
 	\qquad s < 0;\\
	V_{s}(0) &:=& \label{limiteNULLatNULL}
	+\infty\,               , 	\qquad\, s \geq 0.
\end{eqnarray}
 With these stipulations \Ref{aveRIESZpairENERGY} is a lower-semicontinuous function of two of the three angles,
say $(\alpha,\beta)\in [0,\pi]^2\cap\{\alpha+\beta\leq \pi\}$ (and with $\gamma=\pi-\alpha-\beta$), so 
a minimizer for \Ref{MINaveRIESZpairENERGY} always exists.
 It coincides with the minimizer of the total Riesz $s$-energy as defined in \cite{Sch2016arXiv}, see the introduction.

 We will prove the following theorems. 
 The first one lists all the absolute minimizers (see Appendix 1 of \cite{NBK}).
\begin{theorem}\label{absMIN}
 Set $s_3 := \ln(4/9)/\ln(4/3)$. 
 Then for $s\neq s_3$ the optimal Riesz $s$-energy $N=3$ arrangement on $\Sset^1$ is unique (up to rotation). 
 For $s<s_3$ it is given by the antipodal arrangement $(\alpha,\beta,\gamma)=(\frac{\pi}{2},\frac{\pi}{2},0)$ 
(up to permutation), and for $s_3 < s$ by the equilateral configuration 
$(\alpha,\beta,\gamma)=(\frac{\pi}{3},\frac{\pi}{3},\frac{\pi}{3})$.
 At $s=s_3$ both have the same (standardized) Riesz $s_3$-energy (averaged over the pairs.) 
\end{theorem}

 The next theorem lists all absolute maximizers on $\Sset^1$.
\begin{theorem}\label{absMAX}
 For all $s<0$ the completely degenerate triangular configuration (i.e. the one-point arrangement) given by
$(\alpha,\beta,\gamma)= (0,0,\pi)$ (up to permutation) is the unique (up to rotation) absolute maximizer
of \Ref{aveRIESZpairENERGY}.
\end{theorem}

 Recalling the usual notion of a relative minimizer / maximizer of (the extended) \Ref{aveRIESZpairENERGY}, 
we define a \emph{relative minimizer} as any triple $(\alpha_*,\beta_*,\gamma_*)$ of non-negative angles satisfying 
$\alpha_* + \beta_* + \gamma_* = \pi$ 
such that $\langle V_s\rangle (\alpha_*,\beta_*,\gamma_*) \leq \langle V_s\rangle (\alpha,\beta,\gamma)$ for all
triples $(\alpha,\beta,\gamma)$ of non-negative angles satisfying $\alpha + \beta + \gamma = \pi$ in a small enough
neighborhood of $(\alpha_*,\beta_*,\gamma_*)$, with strict ``$<$'' holding for some of them; a \emph{relative maximizer} 
is defined by replacing ``$\leq$'' with ``$\geq$'' and ``$<$'' by ``$>$.''
 
 The next theorem lists all relative minimizers / maximizers on $\Sset^1$ which are not absolute.
\begin{theorem}\label{rel}
  For $s_3< s<-2$ the antipodal arrangement $(\alpha,\beta,\gamma)=(\frac{\pi}{2},\frac{\pi}{2},0)$ (up to permutation) is a
relative minimizer which is not absolute.

 The equilateral configuration $(\alpha,\beta,\gamma)=(\frac{\pi}{3},\frac{\pi}{3},\frac{\pi}{3})$ is a
relative maximizer of \Ref{aveRIESZpairENERGY} for $s<-4$, 
and a relative minimizer for $-4<s<s_3$, neither of which is absolute.
\end{theorem}

 The theorems stated above do not depend on the notion of proper force equilibrium, introduced next in terms of $\alpha$ and $\beta$
derivatives of $\langle V_s\rangle$.
 A triple $(\alpha_*,\beta_*,\gamma_*)$ of non-negative angles satisfying $\alpha_* + \beta_* + \gamma_* = \pi$ is
called a \emph{proper force equilibrium} if both 
$\partial_\alpha \langle V_s\rangle (\alpha,\beta,\pi-\alpha-\beta)|_* = 0$
and
$\partial_\beta \langle V_s\rangle (\alpha,\beta,\pi-\alpha-\beta)|_* = 0$, 
where ``$...|_*$'' means the expression $...$ to its left is evaluated with $\alpha=\alpha_*$ and $\beta=\beta_*$, so 
also $\gamma=\gamma_*$.
 The next theorem lists all proper  equilibria under a repulsive power-law force.
\begin{theorem}\label{prf}
 The one-point arrangement given by
$(\alpha,\beta,\gamma)= (0,0,\pi)$, and the antipodal arrangement given by
$(\alpha,\beta,\gamma)=(\frac{\pi}{2},\frac{\pi}{2},0)$, are 
proper force equilibria for all $s<-1$. 

 The equilateral configuration $(\alpha,\beta,\gamma)=(\frac{\pi}{3},\frac{\pi}{3},\frac{\pi}{3})$ is a
proper repulsive power-law force equilibrium for all $s\in\Rset$. 

 In addition to these $s$-independent 3-particle force equilibria there exists, for each $s<-2$ except $s=-4$,
a non-universal isosceles triangular proper force equilibrium, i.e. its shape depends on $s$.

 This list exhausts all proper repulsive power-law force equilibria of three particles on $\Sset^1$.
\end{theorem}

 As explained in the introduction, a Cauchy principal-value type definition of a pseudo force can be stipulated
whenever the proper force is not defined.
 This vindicates the following.
\begin{definition}\label{pseudoFisNULL}
 When two particles occupy the same point, and $s\geq - 1$, then
the pseudo force of one particle onto another one vanishes.
\end{definition}

 With the help of this definition it is easy to prove
\begin{theorem}\label{psf}
 Both, the completely degenerate triangular configuration given by
$(\alpha,\beta,\gamma)= (0,0,\pi)$ (up to permutation), and the antipodal arrangement given by
$(\alpha,\beta,\gamma)=(\frac{\pi}{2},\frac{\pi}{2},0)$ (up to permutation), are 
non-proper pseudo force equilibria for all $s\geq -1$. 
 These are the only non-proper pseudo force equilibria of $N=3$ point particles on $\Sset^1$.
\end{theorem}

With the help of the notions of a proper, respectively a non-proper pseudo force equilibrium 
we are now in the position to define ``saddle points.''
\begin{definition}
 Any (proper, or non-proper pseudo) force $N$-particle equilibrium which is neither a relative minimum or
relative maximum of the Riesz $s$-energy of an $N$-particle arrangement is called a \emph{saddle point}.
\end{definition}
The next theorem lists all saddle points of \Ref{aveRIESZpairENERGY}.
\begin{theorem}\label{sad}
 The closure of the two families of proper isosceles triangular configurations for $\{s < -2\}\cap\{s\neq-4\}$ 
are saddle points on $\Sset^1$. 
 This includes the equilateral triangular configuration at $s=-4$ and the antipodal arrangement at $s=-2$, and
in a limiting sense, the right triangular configuration at ``$s=-\infty$.''

 The antipodal arrangement (a degenerate isosceles configuration) is a saddle point for all $-2 \leq s <0$.
\end{theorem}
\begin{remark}\label{saddlesECTextended}
 For $s\geq 0$ the degenerate configurations have $\langle V_s\rangle (\alpha,\beta,\gamma)=+\infty$,
and while we can count them as non-proper pseudo Riesz $s$-energy equilibria, a classification into 
relative maximizers / minimizers, or saddles, would be
ambiguous, for any degenerate configuration would have $+\infty$ energy.
 However, if desired, by a compressed compactification (e.g., replacing the extended $V_s(r)$ by $\tanh V_s(r)$ (with
$\tanh(\pm\infty):=\pm 1$), we can assign any coincident pair of particles the compressed (standardized) Riesz pair energy
 $=1$, and then the classification of Theorem \ref{sad} continues to hold for all $s\geq 0$.
\end{remark}

 Lastly, we describe in some detail the non-universal, isosceles, proper force equilibria,
stipulating a representation $(\alpha,\beta,\gamma)(s)$ with $\alpha=\beta$. 
\begin{theorem}\label{iso}
 The family of isosceles triangular proper force equilibria for $s\in(-\infty,-4)$
interpolates continuously and monotonically between a right triangular configuration ($\gamma=\pi/2$), to 
which it converges when $s\downarrow-\infty$, and the equilateral configuration ($\gamma=\pi/3$), to 
which it converges when $s\uparrow -4$. 
 The family of isosceles triangular proper force equilibria for $s\in(-4,-2)$
interpolates continuously and monotonically between the equilateral configuration ($\gamma=\pi/3$), to 
which it converges when $s\uparrow -4$, and the antipodal arrangement ($\gamma=0$), to 
which it converges when $s\uparrow -2$. 

 The asymptotics of its angle $\gamma$ as function of $s$ is given by the following:

\noindent
(a) in a left neighborhood of $\gamma=\pi/2$ (as $s\downarrow -\infty$),
\begin{equation}\label{asympFORsTOminINFTY}
\gamma(s) \asymp \textstyle 
\frac{\pi}{2} - \surd{2}^{1+s},
\end{equation}

\noindent
(b) in a neighborhood of $\gamma=\pi/3$ (for $s\approx -4$),
\begin{equation}\label{asympFORsTOminFOUR}
\gamma(s) =\textstyle \frac{\pi}{3} - \frac{1}{2\surd3} (4+s) + \mathcal{O}((s+4)^2)
\end{equation}

\noindent
(c) in a right neighborhood of $\gamma=0$ (as $s\uparrow -2$),
\begin{equation}\label{asympFORsTOminTWO}
\gamma(s) \asymp 0+ 2^{\tfrac{1}{2+s}}.\qquad
\end{equation}
\end{theorem}

 Our results are illustrated by the following diagram.
\begin{figure}[H]
\centering
\includegraphics[scale=.35]{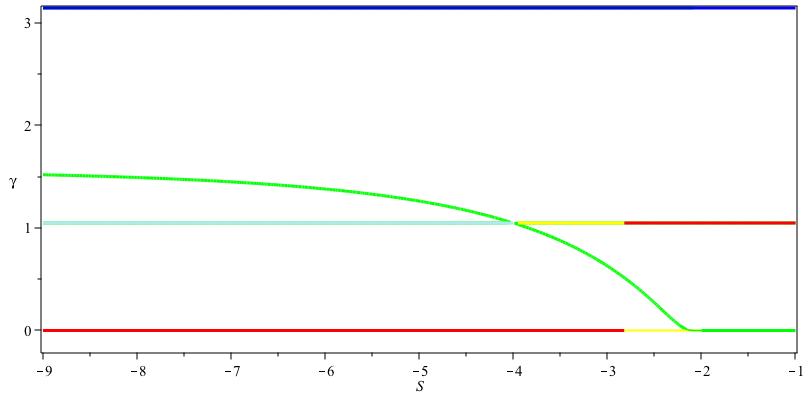} \vspace{-10pt}
\caption{Bifurcation diagram ($\gamma$ vs. $s$) of repulsive power-law force equilibria.\hspace{2truecm}
\centerline{ Color code: 
\MARKED{red}{absolute min.},
\MARKED{yellow}{relative min.}, 
\MARKED{green}{saddle},
\MARKED{BlueGreen}{relative max.}, 
\MARKED{blue}{absolute max}.}}
\label{BIFdiagram}
\end{figure}

\newpage

	\section{Proofs}\label{proofs}\vspace{-10pt}

 We begin with the pseudo force equilibria by proving \vspace{-5pt}
\begin{proposition}\label{equilISisoscPSEUDO}
 All non-proper pseudo force equilibria of three particles on $\Sset^1$ are reflection-symmetric about a 
diameter of $\Sset^1$.\vspace{-5pt}
\end{proposition}

\noindent
\textit{Proof of Proposition \ref{equilISisoscPSEUDO}:} 

 The force of one point particle on another is not proper if and only if $s\geq -1$ and 
both particles occupy the same location. 
 By Definition \ref{pseudoFisNULL}, the pseudo force of any one of these two particles on the other
vanishes. 
 Therefore, when $s\geq -1$ we have the following two possibilities:

(i) all three particles occupy the same position --- in that case the total pseudo force on any one of them 
vanishes, so this arrangement is a pseudo force equilibrium;

(ii) only two of the three particles occupy the same position --- in that case the total pseudo force on any 
one of the two particles with coincidental positions is given by the proper force of the remaining third particle 
on it, while the total force on the third particle is the sum of the proper forces of the two ``coincident'' 
particles on it.
 But then, the only way the total force on each and any one of the three particles can point radial is when the arrangement 
is antipodal, so the antipodal arrangement is a pseudo force equilibrium, too.

Since these two cases exhaust all possibilities, and since in either case the arrangement is manifestly reflection-symmetric
about a diameter of $\Sset^1$, Proposition \ref{equilISisoscPSEUDO} is proved. \hfill $\square$
\smallskip

 The proof of Proposition \ref{equilISisoscPSEUDO} identifies the antipodal
and the single-point arrangements as the only possible $N=3$ pseudo force equilibria, and
thereby also proves Theorem \ref{psf}.

 We now turn our attention to the proper force equilibria. 
 Proposition \ref{psf} has the following counterpart.\vspace{-5pt}
\begin{proposition}\label{equilISisosc}
 All proper repulsive power-law force equilibria of three particles on $\Sset^1$ are reflection-symmetric about a diameter of $\Sset^1$.\vspace{-5pt}
\end{proposition}


\noindent
\textit{Proof of Proposition \ref{equilISisosc}:} 

 The condition for a triple $(\alpha_*,\beta_*,\gamma_*)$ of non-negative angles satisfying $\alpha_* + \beta_* + \gamma_* = \pi$ 
to be a proper force equilibrium of three particles on $\Sset^1$ consists of the pair of equations
$\partial_\alpha \langle V_s\rangle (\alpha,\beta,\pi-\alpha-\beta)|_* = 0$
and
$\partial_\beta \langle V_s\rangle (\alpha,\beta,\pi-\alpha-\beta)|_* = 0$, 
where ``$...|_*$'' means the expression $...$ to its left is evaluated with $\alpha=\alpha_*$ and $\beta=\beta_*$, so 
also $\gamma=\gamma_*$.
 Carrying out the differentiations, and cancelling common factors, yields the pair of equations
\begin{eqnarray} \label{equilA}
   \frac{\cos(\alpha_*)}{\sin^{s+1}(\alpha_*)} + \frac{\cos(\alpha_*+\beta_*)}{\sin^{s+1}(\alpha_*+\beta_*)} =  0,\\
 \label{equilB}
   \frac{\cos(\beta_*)}{\sin^{s+1}(\beta_*)} + \frac{\cos(\alpha_*+\beta_*)}{\sin^{s+1}(\alpha_*+\beta_*)} = 0.
\end{eqnarray}	
 This is equivalent to 
\begin{equation} \label{equilC}
  \frac{\cos(\alpha_*)}{\sin^{s+1}(\alpha_*)} = \frac{\cos(\beta_*)}{\sin^{s+1}(\beta_*)} = \frac{\cos(\gamma_*)}{\sin^{s+1}(\gamma_*)} 
\end{equation}	
together with $\alpha_* + \beta_* + \gamma_* = \pi$, which for all $s\in\Rset$ has the obvious solution 
$\alpha_* = \beta_* = \gamma_* = \pi/3$, namely the equilateral triangular configuration on $\Sset^1$ 
(which is reflection-symmetric through even three different diameters of $\Sset^1$).

\begin{figure}[H]
\centering
\includegraphics[scale=.5]{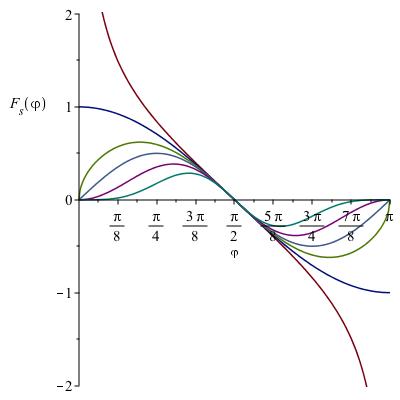} \vspace{-10pt} %
\caption{$F_s(\varphi)$ for $s\in\{-5,-3,-2,-1.5,-1,-0.5\}$. The graphs are reflect\-ion-antisymmetric
about $\varphi=\frac{\pi}{2}$, and $s\mapsto F_s(\varphi)$ is increasing 
when $\varphi\in(0,\frac{\pi}{2})$.}\vspace{-10pt}
\label{FsOFphi}
\end{figure}

 To show that any other solution of \Ref{equilC} satisfying $\alpha_* + \beta_* + \gamma_* = \pi$ is reflection-symmetric
about a diameter of $\Sset^1$, 
 we need to discuss the function $\frac{\cos(\varphi)}{\sin^{s+1}(\varphi)}=:F_s(\varphi)$ for $\varphi\in(0,\pi)$, 
together with its behavior as $\varphi\downarrow 0$ and $\varphi \uparrow \pi$; see Fig.\ref{FsOFphi}.

 First of all, since $\sin(\varphi)\neq 0$ on the open interval $(0,\pi)$, for each $s\in\Rset$
the function $F_s(\varphi)$ is infinitely often continuously differentiable on $(0,\pi)$, 
and since $\sin(\varphi)=\sin(\pi-\varphi)$
and $\cos(\varphi)=-\cos(\pi-\varphi)$, we have $F_s(\varphi)=-F_s(\pi-\varphi)$ 
on $\varphi\in(0,\pi)$, i.e. $F_s(\varphi)$ is reflection-antisymmetric about $\varphi=\pi/2$, 
where it vanishes. 
 Moreover, $F_s(\varphi)>0$ on $(0,\pi/2)$ and, by antisymmetry, $F_s(\varphi)<0$ on $(\pi/2,\pi)$.
 As a consequence, any triple $(\alpha_*,\beta_*,\gamma_*)$ of non-negative angles satisfying \Ref{equilC} and
 $\alpha_* + \beta_* + \gamma_* = \pi$ must have $F_s(\alpha_*)\geq 0$.
 For suppose not, i.e. suppose $F_s(\alpha_*)<0$. 
 Then $(\alpha_*,\beta_*,\gamma_*)\in(\pi/2,\pi]^3$, which violates the constraint
$\alpha_* + \beta_* + \gamma_* = \pi$.

 Thus it suffices to discuss $F_s$ restricted to the pre-image of $F_s\geq 0$.
 Since the behavior of $F_s(\varphi)$ near $\varphi=0$ depends on $s$, we now distinguish the 
cases $s>-1$, $s=1$, and $s<-1$.

{\bf{Case}} $s>-1$. 
 Since $\sin(0)= 0 =\sin(\pi)$, while $\cos(0) = 1 = -\cos(\pi)$, when $s>-1$
the function $\varphi\mapsto F_s(\varphi)$ is not defined at $\varphi=0$ and not at $\varphi=\pi$, 
where it diverges to $\pm\infty$, respectively.
 This already rules out any degenerate configuration (at least one angle $=0$) from being a proper force
equilibrium when $s>-1$. 

 Moreover, it is easy to see that $F_s(\varphi)$ is strictly monotonically decreasing on  $\varphi\in(0,\pi)$,
when $s>-1$. 
 For any $\ell\geq 0$, the equation $F_s(\varphi)=\ell$ therefore has only one solution $\varphi_*\in(0,\pi)$.
 But this means that $\alpha_* = \beta_* = \gamma_* $, and now $\alpha_* + \beta_* + \gamma_* = \pi$ forces $\alpha_*=\pi/3$.

 In geometric terms, when $s>-1$ the equilateral triangular configuration is the only proper force equilibrium 
of three point particles on $\Sset^1$.

{\bf{Case}} $s=-1$. 
 Clearly, $F_{-1}(\varphi)\equiv \cos(\varphi)$, so $F_{-1}(\varphi)$ is well-defined on the closed interval $[0,\pi]$.
 Yet, since $\cos(\varphi)$ decreases strictly monotonically from $1$ to $-1$ when $\varphi$ varies through $[0,\pi]$, 
the same conclusion as in the previous case holds: the equilateral triangular configuration is the only proper
$s=-1$-force equilibrium of three point particles on $\Sset^1$.

{\bf{Case}} $s<-1$. 
 Now $F_s(\varphi)$ is well-defined on the closed interval $\varphi\in[0,\pi]$, with $F_s(0) = 0 =F_s(\pi)$.
 Since also $F_s(\pi/2)=0$, the level value $\ell=0$ now offers three different possible values for the three
angles $\alpha_*$, $\beta_*$, $\gamma_*$ to satisfy \Ref{equilC} with  $\alpha_* + \beta_* + \gamma_* = \pi$.
 Up to permutations, there are exactly two possibilities: 

(i)\ $(\alpha_*,\beta_*,\gamma_*) = (0,0,\pi)$ (the completely degenerate configuration);

(ii) $(\alpha_*,\beta_*,\gamma_*) = (\pi/2,\pi/2,0)$ (the antipodal arrangement).
 
\noindent
 This already establishes the two degenerate isosceles triangular configurations as proper force equilibria for
all $s<-1$.
 Obviously they are reflection-symmetric about some diameter of $\Sset^1$.

 Next, let $\ell > 0$. 
 In this case there is no possible $\varphi_* \in[\pi/2,\pi]$ for which $F_s(\varphi_*)=\ell$, and we seek $\varphi_*\in(0,\pi/2)$
for which $F_s(\varphi_*)=\ell$.
 We show that, depending on $\ell$, there can be exactly two, exactly one, or no value of $\varphi_*\in(0,\pi/2)$ 
for which $F_s(\varphi_*)=\ell$.

 From the facts that $F_s(0) = 0 =F_s(\pi/2)$ when $s<-1$, 
that $F_s(\varphi)>0$  for $\varphi\in (0,\pi/2)$, and that $\varphi\mapsto F_s(\varphi)$
is continuous, it follows that $F_s(\varphi)$ has a maximum in $(0,\pi/2)$. 
 Clearly, if $\ell$ surpasses this maximum value of $F_s(\varphi)$, then there is no $\varphi_*$ satisfying $F_s(\varphi_*)=\ell$.

 Since $\varphi\mapsto F_s(\varphi)$ is differentiable (infinitely often, actually) on $(0,\pi/2)$, any such maximum is taken at a 
critical point, where $F_s^\prime(\varphi)=0$.
 Differentiating $F_s$ w.r.t. $\varphi$ and simplifying, we find that the equation $F_s^\prime(\varphi)=0$ for $\varphi\in(0,\pi/2)$
is equivalent to
\begin{equation} \label{equilD}
\sin^2\varphi + (s+1)\cos^2\varphi = 0; \qquad \varphi \in (0,\pi/2),
\end{equation}	 
which in turn (using $\sin^2\varphi + \cos^2\varphi =1$) is equivalent to 
\begin{equation} \label{equilE}
1 + s \cos(\varphi)^2 = 0; \qquad \varphi \in (0,\pi/2).
\end{equation}	 
 Since $s<-1$, there is a unique $\phi = \arccos \sqrt{-1/s} \in (0,\pi/2)$ satisfying \Ref{equilE}.

 Thus, for $s<-1$ there exists in $(0,\pi/2)$ a unique maximum of $F_s(\varphi)$, at $\varphi=\phi$, with value $L(s)$ (say).
 So when $s<-1$, then $F_s(\varphi)$ increases monotonically and continuously from $0$ to $L(s)$ when $\varphi$ increases 
through $[0,\phi]$, and $F_s(\varphi)$ decreases monotonically and continuously from $L(s)$ to $0$ when $\varphi$ increases 
through $[\phi,\pi/2]$.

 It follows that if $\ell = L(s)$, then there is a unique value $\varphi_*$ satisfying $F_s(\varphi_*)=\ell$, namely
$\varphi_* = \phi(s)$. 
 If \Ref{equilC} together with $\alpha_* + \beta_* + \gamma_* = \pi$ is satisfied with $\alpha_*=\phi(s)$,  
then all three angles $\alpha_*$, $\beta_*$, $\gamma_*$ must coincide. 
 But this can lead to a proper force equilibrium if and only if $\phi(s) = \pi/3$. 
 Incidentally, by \Ref{equilE} this happens if and only if $1+s/4=0$, or $s=-4$. 
 As already remarked, the equilateral configuration is manifestly reflection-symmetric about a diameter of $\Sset^1$.

 Yet if $0<\ell< L(s)$, there are precisely two different values of $\varphi_*$, satisfying $0<\varphi_*^{(1)}<\varphi_*^{(2)}<\pi/2$,
for which $F_s(\varphi_*)=\ell$.
 This now implies that for any possible solution of \Ref{equilC} with value $F_s(\alpha_*)=\ell\in(0,L(s))$,
satisfying $\alpha_* + \beta_* + \gamma_* = \pi$,
at least two of the three angles $\alpha_*$, $\beta_*$, $\gamma_*$ must coincide, because there are only two possible 
values, $\varphi_*^{(1)},\varphi_*^{(2)}$ which any of the three angles  $\alpha_*$, $\beta_*$, $\gamma_*$ can take.
 But then, since for $\ell>0$ none of the angles can be $0$, 
any possible solution is a proper isosceles triangular configuration with its corners on $\Sset^1$, hence
reflection-symmetric about a diameter of $\Sset^1$.

 The proof of the proposition is complete.\hfill $\square$
\medskip

  Proposition \ref{equilISisosc}, and more so its proof, reveal a number of important facts about 
the proper force equilibria of three particles on $\Sset^1$, which we summarize as

\begin{corollary}$\phantom{nix}$

\smallskip\noindent
{\bf{A}}. All these proper force equilibria are isosceles
triangular arrangements, to which we count the proper isosceles triangular configurations on $\Sset^1$ as well as their 
degenerate limits,\footnote{We remark that the degenerate linear configurations on $\Sset^1$ with one particle at one
  end point and two at the other end point of a secant which is not a diameter are \emph{not} degenerate limits of
  proper isosceles triangular configurations on $\Sset^1$.
   While the antipodal arrangement is represented by $(\alpha,\beta,\gamma)=(\pi/2,\pi/2,0)$ (up to permutation), 
   a linear but non-antipodal arrangement is the limit of a family of non-isosceles proper triangles with corners
   on $\Sset^1$ which leads to a representation $(\alpha,\beta,\gamma)=(\alpha,\pi-\alpha,0)$ (up to permutation), 
   with $\alpha\neq \pi/2$.}
the antipodal arrangement and the completely degenerate configuration (the single-point arrangement).

\smallskip\noindent
{\bf{B}}. The equilateral triangular configuration is a universal proper force equilibrium.
  When $s\geq -1$, it is the only proper force equilibrium.

\smallskip\noindent
{\bf{C}}. The antipodal and the one-point arrangements are proper force equilibria for $s<-1$.
  These are the only degenerate proper force equilibria. 
\end{corollary}

 Furthermore, Propositions \ref{equilISisoscPSEUDO} and \ref{equilISisosc}, and their proofs, in concert reveal 
\begin{corollary}$\phantom{nix}$

\smallskip\noindent
{\bf{D}}. The antipodal and the one-point arrangements are universal force equilibria (proper for $s<-1$, 
pseudo for $s\geq -1$).  

\smallskip\noindent
{\bf{E}}. There are no pseudo force equilibria other than those just listed; in particular,
there are no non-universal pseudo equilibria --- any non-universal force equilibrium is necessarily a proper 
force equilibrium. 

\smallskip\noindent
{\bf{F}}.  Non-universal force equilibria can at most exist when $s<-1$.
\end{corollary}

 These corollaries show that Propositions \ref{equilISisoscPSEUDO} and \ref{equilISisosc}, and their proofs, 
exhaust the list of force equilibria (proper or pseudo) of  $N=3$ particles on $\Sset^1$ when $s\geq -1$:
the equilateral configuration (proper), 
the antipodal arrangement (pseudo), 
and the single-point arrangement (pseudo). 
 Their proofs have also yielded, as a byproduct, that these three equilibria (proper for $s<-1$) are universal.
 Since a universal equilibrium is by definition $s$-independent, and since the ones we have identified also exhaust the 
list of all equilibria for when $s\geq -1$, it follows that our  list of universal equilibria is complete. 

 Propositions \ref{equilISisoscPSEUDO} and \ref{equilISisosc} and their proofs do not identify any 
non-universal equilibria, which by Corollary 2  can at most exist when $s<-1$.
 This is the content of the next proposition.
\begin{proposition}\label{equilISO}
There are exactly two continuous families of non-universal proper equilibria, one for $s\in(-\infty,-4)$, the other
for $s\in(-4,-2)$.
\end{proposition}

\smallskip
\noindent
\textit{Proof of Proposition \ref{equilISO}:} 

 We already showed that there are no non-universal equilibria for $s\in[-1,\infty)$, and since pseudo force
equilibria by definition cannot exist for $s<-1$, any non-universal equilibrium 
must be a proper force equilibrium with $s<-1$. 

 Thanks to Proposition \ref{equilISisosc}, any proper equilibrium for $s<-1$ must correspond to a point on a height of
the fundamental triangle. 
 Its three heights meet at the center of the fundamental triangle, and we already know that the center corresponds to the
equilateral equilibrium configuration. 
 Moreover we know that the two end points of each height also correspond to equilibra, namely the antipodal
arrangement (where a height meets the midpoint of a side) and the completely degenerate configuration (where a height 
meets a corner). 
 These equilibria are universal. 
 Thus any non-universal equilibrium would have to correspond to a point on the height either between the center and a midpoint
of a side, or between the center and a corner, of the fundamental triangle.
 
\begin{figure}[H]
\centering
\includegraphics[scale=.5]{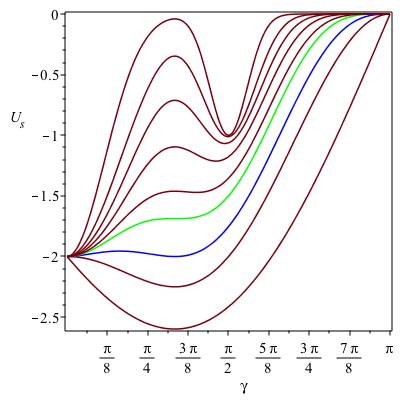} \vspace{-10pt} 
\caption{$U_s(\gamma)$ for $s\in\{-30,-15,-10,-7,-5,\MARKED{green}{-4},
\MARKED{blue}{ \frac{\ln(4/9)}{\ln(4/3)}},-2,-1\}$.}\vspace{-10pt}
\label{UsGRAPHS}
\end{figure}

 Since each height represents the same geometrical situation, we work with the height which corresponds to choosing $\alpha=\beta$, 
and parametrize the height with $\gamma$. 
 The equilibria are then determined by the critical points 
$\gamma_*$ of the map $\gamma\mapsto \langle V_s\rangle \big(\tfrac{\pi-\gamma}{2},\tfrac{\pi-\gamma}{2},\gamma\big)$,
the standardized Riesz $s$-energy averaged over the pairs of an isosceles triangle 
$\big(\frac{\pi-\gamma}{2},\frac{\pi-\gamma}{2},\gamma\big)$ with $\gamma\in[0,\pi]$. 
 For $s<-1$ these agree with the critical points of the map $\gamma\mapsto U_s(\gamma)$, given by
\begin{equation}\label{UsOFgamma}
U_s (\gamma) : =  
- {2}\sin^{|s|}(\tfrac{\pi-\gamma}{2}) -  \sin^{|s|}(\gamma) ; \qquad s< 0.
\end{equation}	 
 Fig.\ref{UsGRAPHS} shows $U_s$ graphed vs. $\gamma$ for a selection of $s$ values.
 The graphs are monotonically ordered with $s$, decreasing with $s$, which follows from \Ref{UsOFgamma}.
 Even though the figure displays the graphs only for a finite selection of $s$-values, it
guides us toward our goal: it shows the $s$-independent critical points $\gamma_*\in\{0,\pi/3,\pi\}$ for $s<-1$,
and it suggests exactly one additional critical point $\gamma_*(s)\in(0,\pi/3)$ when $s\in (-4,-2)$, and
exactly one additional critical point $\gamma_*(s)\in(\pi/3,\pi/2)$ when $s\in(-\infty,-4)$. 
 Thus guided by Fig.\ref{UsGRAPHS} we show:

\medskip
\noindent
(a) the three universal ones are the only equilibria if $-2\leq s <-1$ or $s=-4$;

\noindent
(b) $\forall\; s\in(-\infty,-4)\cup(-4,-2)$ there is exactly one non-universal equilibrium.
\smallskip

 Recall that for $s<-1$ all force equilibria are proper.
 Hence taking the $\gamma$-derivative of $U_s(\gamma)$ (see \Ref{UsOFgamma}), using 
$\sin(\tfrac{\pi-\gamma}{2}) = \cos(\tfrac{\gamma}{2})$, which yields
\begin{equation}\label{UsOFgammaPRIME}
U_s^\prime (\gamma)  = |s|\Big(\cos^{|s|-1}(\tfrac{\gamma}{2})\sin(\tfrac{\gamma}{2})  -  \sin^{|s|-1}(\gamma)\cos(\gamma)\Big) ,
\end{equation}	 
we look for zeros of \Ref{UsOFgammaPRIME} for $\gamma\in(0,\pi/2)$ other than $\gamma_*=\pi/3$.
 Recall that we already showed that all non-degenerate configurations have their angle $\gamma_*(s)\in(0,\pi/2)$.
 Using  $\sin(\gamma) = 2\sin(\tfrac{\gamma}{2})\cos(\tfrac{\gamma}{2})$ and $\cos(\gamma) = 1 - 2\sin^2(\tfrac{\gamma}{2})$,
and simplifying, we rewrite \Ref{UsOFgammaPRIME} as
\begin{equation}\label{UsOFgammaPRIMEagain}
U_s^\prime (\gamma)  = |s|\cos^{|s|-1}(\tfrac{\gamma}{2})\sin(\tfrac{\gamma}{2})
\big[1  -  2^{|s|-1}\sin^{|s|-2}(\tfrac{\gamma}{2})\big(1 -2 \sin^2(\tfrac{\gamma}{2}\big)\big] .
\end{equation}	 
 For $\gamma\in(0,\pi)$, both $\cos(\tfrac{\gamma}{2})>0$ and $\sin(\tfrac{\gamma}{2})>0$, so a zero of $U_s^\prime(\gamma)$ 
for $\gamma\in(0,\pi)$ can only come from the factor in square parentheses.
\newpage

 Defining $2\sin(\tfrac{\gamma}{2})=:\xi$ the factor in square parentheses becomes 
\begin{equation}\label{UsOFgammaPRIMEfactorXI}
\xi^{|s|} -  2\xi^{|s|-2}  + 1  =: g_s(\xi),
\end{equation}	 
and we seek its zeros $\xi_*$ for $\xi\in (0,\surd{2})$; note that
$0<\gamma_*(s)<\pi/2$ for $s<-2$ implies that $0<\xi_*(s)<\sqrt{2}$ for $s<-2$.

 Obviously $\xi_*^{(1)} = 1$ is one zero of $g_s(\xi)$, independently of $s$ 
--- it corresponds to $\gamma = \pi/3$, the equilateral triangular configuration. 
 We now ask whether there are any other zeros of $g_s(\xi)$ in $(0,\surd{2})$.

 To answer that question we compute 
$g_s^\prime(\xi) = |s|\xi^{|s|-1}  -  2(|s|-2)\xi^{|s|-3} = |s|\xi^{|s|-3} \big(\xi^{2}  -  2\tfrac{|s|-2}{|s|}\big)$.
 Clearly, the factor $|s|\xi^{|s|-3}>0$ when $\xi>0$.

 Turning first to claim (a), part one, let $-2 \leq s < -1$. 
 Then the factor $\big(\xi^{2}  -  2\tfrac{|s|-2}{|s|}\big)>0$, too. 
 Thus, if $-2 \leq s < -1$ then $\xi\mapsto g_s(\xi)$ is strictly monotonic increasing on $(0,\surd{2})$, and since $\xi=1$ is
a (universal) zero of $g_s(\xi)$, there are no other zeros.
 This establishes the first part of claim (a).
 On the other hand, let $s=-4$; then, since $g_{-4}(\xi) = \xi^4 - 2\xi^2 +1 = (\xi^2 -1)^2$, once again $\xi_*^{(1)}=1$ is 
the only zero of $g_s(\xi)$ in the open interval $(0,\surd{2})$. 
 This establishes the second part of claim (a).

\begin{remark}
 We note that the universal zero $\xi_*^{(1)}=1$ of $g_s(\xi)$ is of degree 1 for all 
$s\in\Rset$ except $s=-4$, when it is of degree 2.
 This will play a role when we discuss the ``monkey saddle.'' 
\end{remark}

 We turn to proving claim (b). 
 Let  $s\in(-\infty,-2)$, but $s\neq -4$.
 We know that $g_s(1) =0$. 
 We also have $g_s^\prime(1) = s+4 \neq 0$ (for $s\neq -4$). 
 So $g_s(\xi)$ changes its sign at its zero $\xi_*^{(1)}=1$. 
 But now we have $g_s(0) = 1 = g_s(\surd{2})$.
 Therefore, $g_s(\xi)$ has to change its sign at least twice on $(0,\surd{2})$, and therefore $g_s(\xi)$ has at least
two distinct zeros, $\xi_*^{(1)}=1$ and $\xi_*^{(2)}\equiv \xi_*(s)\neq 1$, in $(0,\surd{2})$. 
 We remark that  $\xi_*(s)\neq 1$ must depend on $s$, since $s\mapsto \xi^{|s|}-2\xi^{|s|-2}$ is not constant for 
$\xi\in(0,\surd{2})$ unless $\xi=1$. 

 We show that there are no other zeros of $g_s(\xi)$ in this open interval. 
 This follows by noting that for $s< -2$, the derivative function $g_s^\prime(\xi)$ computed above 
has a unique zero on $(0,2)$, namely at $\xi_\circ(s) = \sqrt{2-{4}/{|s|}}\in (0,\surd{2})$.
 And so, since $g_s(0)=1=g_s(\surd{2})$, it follows that
$g_s(\xi)$ has a unique minimum at $\xi_\circ(s) \in (0,\surd{2})$, 
 establishing that it cannot have more than 2 distinct zeros in $(0,\surd{2})$. 
 This proves claim (b).

 Having shown that there is precisely one non-universal equilibrium for each $s\in(-\infty,-2)$, with the exception
of $s=-4$, the fact that these form two continuous families, one for $s\in(\infty,-4)$, 
the other for $s\in(-4,-2)$, now follows from the implicit function theorem. 
 These families are disconnected because all equilibria at $s=-4$ are universal.
 Indeed, the implicit function theorem fails to apply at $s=-4$, because $\xi_*(-4)=1$ and 
$g_{-4}^\prime(1)=0$. 

Proposition \ref{equilISO} is proved. \hfill $\square$
 \smallskip

 Proposition \ref{equilISO} together with the proof of Proposition \ref{equilISisosc} proves Theorem \ref{prf}.
 \smallskip

 We are now ready for the 

\smallskip
\noindent
\textit{Proof of Theorem \ref{iso}:}

 We prove monotonic decrease of $s\mapsto\gamma_*(s)$ by differentiating $g_s(\xi_*(s))\equiv 0$ w.r.t. $s$,
viz.
\begin{equation}\label{UsOFgammaPRIMEfactorXIderiv}
0 = \tfrac{\dd{}}{\dd{s}} g_s(\xi_*(s)) = 
 - \big( \xi_*^{2}(s) -2\big)\xi_*^{|s|-2}(s) \ln \xi_*(s)  + g_s^\prime(\xi_*(s)) \xi_*^\prime(s).
\end{equation}	 
 Thus, for $s\neq -4$, we find
\begin{equation}\label{UsOFgammaPRIMEfactorXIderivREWRITE}
 \xi_*^\prime(s)
=
 \frac{\big( \xi_*^{2}(s) -2\big)\xi_*^{|s|-2}(s) \ln \xi_*(s)}{  g_s^\prime(\xi_*(s))}.
\end{equation}	 
 Since $0<\xi_*(s)<\sqrt{2}$ for $s<-2$, we have $\xi_*^{2}(s) -2< 0$ for $s<-2$.
 Of course, $\xi_*(s)>0$ implies $\xi_*^{|s|-2}(s)>0$. 
 Moreover, $\ln \xi_*(s)>0$ for $\xi_*(s)>1$, which is the case when $s<-4$, 
and  $\ln \xi_*(s)<0$ for $\xi_*(s)<1$,  which is the case when $-4<s<-2$.
 We now show that $g_s^\prime(\xi_*(s))>0$ when $s<-4$, 
and $g_s^\prime(\xi_*(s))<0$ when $-4<s<-2$.
 This implies that $\xi_*^\prime(s)<0$ for all $s\in\!(-\infty,-4)\cup(-4,-2)$, proving monotonic decrease as claimed in
Theorem~\ref{iso}. 

  As to the sign of $g_s^\prime(\xi_*(s))$, recall that
$\xi_*=1$ is the only universal zero of $g_s(\xi)$ in $(0,\surd2)$, i.e.
$g_s(1)=0$; and recall that $g_s(0) = 1 = g_s(\surd2)$.  
  And so, since $g_s^\prime(1)=s+4$, we have that $g_s^\prime(1)>0$ for $s\in(-4,-2)$, and $g_s^\prime(1)<0$ for $s\in(-\infty,-4)$.
  Therefore, the non-universal zero $\xi_*(s)\in(0,\surd2)$ of $g_s(\xi)$ is in $(0,1)$ for $s\in(-4,-2)$, and 
in $(1,\surd2)$ for $s\in(-\infty,-4)$.
  Therefore  $g_s^\prime(\xi_*(s))>0$ when $s<-4$,  and $g_s^\prime(\xi_*(s))<0$ when $-4<s<-2$, as claimed.
  The proof of the monotonic decrease of $s\mapsto \xi_*(s)$ is complete. 

 We next prove that $\xi_*(s)\downarrow 0$ as $s\uparrow -2$, and that $\xi_*(s)\uparrow \surd2$ as $s\downarrow -\infty$.

 Let first $s\in(-4,-2)$. 
 We already know that for $s\in(-4,-2)$, we have $g_s^\prime(1)>0$ and $g_s^\prime(\xi_*(s))<0$,
so $g_s(\xi)$ has its minimum at $\xi_\circ(s) \in (0,1)$ when $s\in(-4,-2)$. 
 We already computed $\xi_\circ(s) = \sqrt{2-{4}/{|s|}}$. 
 Clearly, since $0<\xi_*(s)<\xi_\circ(s)$ when $s\in(-4,-2)$, and since 
$\sqrt{2-{4}/{|s|}}\downarrow 0$ when $s\uparrow -2$, it follows that $\xi_*(s)\downarrow 0$ when $s\uparrow-2$.
 This is equivalent to $\gamma_*(s)\downarrow 0$ when $s\uparrow-2$.

 Let now  $s\in(-\infty,-4)$. 
 We reason similarly as above, but now use that $g_s^\prime(1)<0$ and $g_s^\prime(\xi_*(s))>0$ 
for $s\in(-\infty,-4)$, which implies that $g_s(\xi)$ has its minimum 
at $\xi_\circ(s) \in (1,\surd2)$ when $s\in(-\infty,-4)$. 
 Now $\xi_\circ(s) < \xi_*(s)<\surd2$ together with $\xi_\circ(s) \uparrow\surd2$ when $s\downarrow-\infty$ proves
that $\xi_*(s) \uparrow\surd2$ when $s\downarrow-\infty$. 
 This is equivalent to $\gamma_*(s)\uparrow \pi/2$ when $s\downarrow-\infty$.

 Next, we show that $\xi_*(s)\to 1$ if $s\to-4$; equivalently: $\gamma_*(s)\to\pi/3$ when $s\to-4$.
 This is accomplished with the help of standard analytical bifurcation theory. 
 Namely, by inspecting the second $\gamma$-derivative of $U_s(\gamma)$ at $\gamma=\pi/3$ one finds that
the universal equilibrium family $s\mapsto \gamma_* = \pi/3$ yields a relative minimum of $U_s(\gamma)$ for
$s>-4$ and a relative maximum for $s<-4$, while the second $\gamma$-derivative of $U_s(\gamma)$ at $\gamma=\pi/3$ vanishes at 
$s=-4$. 
 Since $(s,\gamma)\mapsto U_s(\gamma)$ is real analytic about $(s,\gamma)=(-4,\pi/3)$, analytical bifurcation theory \cite{Sattinger}
reveals that at $s=-4$ two continuous families of non-universal equilibria branch off of the universal equilateral one. 
 In a neighborhood of $s=-4$ it can be computed by setting $s = - 4 +2\sigma$, $|\sigma|\ll 1$, and 
$\xi_*^2(s) = 1 + \eta(\sigma)$, with $0<|\eta(\sigma)|\ll 1$, and Taylor-expanding \Ref{UsOFgammaPRIMEfactorXI}.
 After obvious cancellations, this yields
\begin{equation}\label{UsOFgammaPRIMEfactorXIexpand}
\sigma\eta + \big(1+\tfrac12\sigma(\sigma-1)\big)\eta^{2} + {\mathcal{O}}(\eta^3) = 0;
\end{equation}	 
since $\eta\neq 0$, we obtain
\begin{equation}\label{UsOFgammaPRIMEfactorXIexpandSOLVE}
\eta + {\mathcal{O}}(\eta^2) = - \frac{2\sigma}{2+\sigma(\sigma-1)} = -\sigma + {\mathcal{O}}(\sigma^2),
\end{equation}	
i.e. to leading order: $\eta = - \sigma$.
 Therefore, to the same order in $\sigma= (4+s)/2$ we have $\xi_*(s) = 1 - \frac14(s+4)+\mathcal{O}((s+4)^2)$. 
 Clearly, the (locally analytic) non-universal equilibrium families which bifurcate off of the universal equilateral 
equilibrium family at $s=-4$ converges to the equilateral equilibrium when $s\to-4$.
 But we already proved that for each $s\neq -4$ (though $s<-2$) there is a unique (up to rotation) non-universal 
equilibrium, our just determined bifurcated equilibrium families must consist precisely of these unique (up to rotation) 
non-universal equilibria.
 Therefore the non-universal equilibrium angle $\gamma_*(s)$ converges to $\pi/3$ as $s\to-4$, as claimed.

 Incidentally, the proof of the convergence $\xi_*(s)\to1$ as $s\to-4$ already establishes the representation of
the non-universal equilibrium $\gamma_*(s)$ near $s=-4$ given by \Ref{asympFORsTOminFOUR}.
 It remains to establish the asymptotic behavior as $s\downarrow -\infty$, and as $s\uparrow-2$, listed in Theorem \ref{iso}. 

 First, as $s\uparrow-2$, we can solve \Ref{UsOFgammaPRIMEfactorXI} by noting that
as $\xi\downarrow 0$, we have $\xi^{|s|}\ll \xi^{|s|-2}\ll 1$, so \Ref{UsOFgammaPRIMEfactorXI} becomes
to leading non-vanishing order in $s$:
\begin{equation}\label{UsOFgammaPRIMEfactorXIsLESSminTWO}
 1  -  2\xi^{|s|-2} +\mathcal{O}(\xi^{|s|}) = 0, 
\end{equation}	 
or 
\begin{equation}\label{UsOFgammaPRIMEfactorXIsLESSminTWOsolve}
\xi^{|s|-2}  +\mathcal{O}(\xi^{|s|})= \tfrac12,
\end{equation}	 
which (to leading order) yields precisely \Ref{asympFORsTOminTWO}.

 Lastly, as $s\downarrow-\infty$, we rewrite \Ref{UsOFgammaPRIMEfactorXI} as
\begin{equation}\label{UsOFgammaPRIMEfactorXIsTOminINFTY}
 1  -  2\xi^{-2} + \xi^{s} = 0
\end{equation}
and to leading non-vanishing order in $s$ find the solution
\begin{equation}\label{UsOFgammaPRIMEfactorXIsTOminINFTYsolve}
\xi_*(s) \asymp \surd2\left(1- \surd2^{s}\right).
\end{equation}
 Setting $\gamma_*(s) = \pi/2 - 2\lambda$ with $|\lambda|\ll 1$, then
using $\xi_*(s)= 2\sin\big(\frac{\gamma_*(s)}{2}\big) = \surd2\big(\cos\lambda-\sin\lambda\big)\asymp \surd2\big(1-\lambda\big)$
yields precisely \Ref{asympFORsTOminINFTY}.

Theorem \ref{iso} is proved. \hfill $\square$
\medskip

 At this point, with Theorems \ref{prf}, \ref{psf}, and \ref{iso} proved, we have rigorously identified 
all possible repulsive power-law force equilibria, proper or pseudo, of $N=3$ point particles on $\Sset^1$. 
 It remains to prove our ``comparison theorems'' about the Riesz $s$-energies of the equilibrium
arrangements, Theorems \ref{absMIN}, \ref{absMAX}, \ref{rel}, and \ref{sad}.
  We will now first prove Theorem \ref{absMIN}, and Theorem \ref{absMAX} en passe,
then Theorem \ref{rel}, and finally Theorem~\ref{sad}.
\begin{remark}
 Fig.\ref{UsGRAPHS}, which up to a $\gamma$-independent shifting and scaling shows the Riesz $s$-energy 
averaged over the particle pairs in an isosceles triangle as a function of $\gamma$, parametrized by about a 
dozen $s\leq -1$, offers guidance for the proof of Theorem \ref{absMIN}.
 Indeed, since we have proved that all equilibria are isosceles triangles, their degenerate limits included, 
it suffices to prove that  $\langle V_s\rangle \big(\tfrac{\pi-\gamma}{2},\tfrac{\pi-\gamma}{2},\gamma\big)$
has the features illustrated (with the help of $U_s(\gamma)$) in Fig.\ref{UsGRAPHS}.
\end{remark}

\smallskip\noindent
\textit{Proof of Theorem \ref{absMIN}:}

 To determine the absolutely minimizing arrangement we need the energies of all the equilibria as functions of $s$.
 We begin with the universal equilibria.

\noindent
 Equilateral:
$\langle V_s\rangle \big(\tfrac{\pi}{3},\tfrac{\pi}{3},\tfrac\pi3\big) =\frac1s\left(\frac{1}{\surd3^s}-1\right)$ $\forall\, s\in\Rset$.

\noindent
 Antipodal:
$\langle V_s\rangle \big(\tfrac{\pi}{2},\tfrac{\pi}{2},0\big) = \frac1s\left(\frac23\frac{1}{2^s}-1\right)$  if $s <0$;
$\langle V_s\rangle \big(\tfrac{\pi}{2},\tfrac{\pi}{2},0\big) = +\infty$ if $s\geq 0$.

\noindent
 Single-point:
$\langle V_s\rangle \big(0,0,\pi\big) = -\frac1s$ if $s <0$;
$\langle V_s\rangle \big(0,0,\pi\big) = +\infty$ if $s\geq 0$.

 Since we have already proved that for $s\geq -2$ there are only the three universal equilibria $\gamma_*\in\{0,\pi/2,\pi_3\}$, 
when $s\geq -2$ we only need to compare their energies to identify the minimizer(s). 
 This is pretty trivial when $s\geq 0$, for then the two degenerate equilibrium configurations have $+\infty$ Riesz $s$-energy, 
and this proves that the equilateral configuration is the absolute minimizer when $s\geq 0$.
 When $-2\leq s < 0$, we manifestly have $3\surd3^{|s|} > 2\surd2^{|s|} >0$, which demonstrates that 
the equilateral configuration is the absolute minimizer also when $-2\leq s< 0$.

 When $s<-2$ also the energies of the non-universal families are needed for the comparison.
 For the non-universal families we do not have an explicit formula, but all we need is a lower estimate.
 We have
\begin{lemma}\label{nonUNIVenergyBOUND}
 $\forall\,s \in (-\infty,-4)\cup (-4,-2):$
$$
\langle V_s\rangle \big(\tfrac{\pi-\gamma_*(s)}{2},\tfrac{\pi-\gamma_*(s)}{2},\gamma_*(s)\big) > 
\langle V_s\rangle \big(\tfrac{\pi}{2},\tfrac{\pi}{2},0\big).
$$

\end{lemma}

\smallskip\noindent
\textit{Proof of Lemma \ref{nonUNIVenergyBOUND}:} 

 We note that the second $\gamma$-derivative $U_s^{\prime\prime}(\gamma)$, of $U_s(\gamma)$ given in \Ref{UsOFgamma},
when evaluated at the critical points $\gamma=\gamma_*$ of $U_s(\gamma)$, determines whether a critical point is a relative 
minimum or maximum --- unless $U_s^{\prime\prime}(\gamma_*)$ vanishes, in which case higher derivatives decide
(as long as they exist).

 For the antipodal arrangement, $\gamma_*=0$, we have $U_s^{\prime\prime}(0)= -{s}/{2}>0$ for $s<-2$, 
so the antipodal arrangement is a local minimizer for all $s<-2$. 

 For the equilateral configuration, $\gamma_*=\frac\pi3$, we have 
$U_s^{\prime\prime}(\frac\pi3) = -\frac{s}{2}(\frac{2}{\surd3})^{s}(4+s)$ for $s<-2$, so
$U_s^{\prime\prime}(\frac\pi3)>0$ for $s>-4$ and
$U_s^{\prime\prime}(\frac\pi3)<0$ for $s<-4$; 
at $s=-4$ the second derivative vanishes, but differentiation shows that $U_s^{\prime\prime\prime}(\pi/3)\neq 0$ for $s=-4$.
 So the equilateral configuration is a saddle at $s=-4$, a relative minimum for $s>-4$ and a relative maximum for $s<-4$.

 For the completely degenerate configuration, $\gamma_*=\pi$, we have $U_s^{\prime\prime}(\pi)\equiv 0$  for $s<-2$, 
yet it is easy to show that it has the highest possible Riesz $s$-energy for all $s<0$:

\smallskip\noindent
\textit{Proof of Theorem \ref{absMAX}:} 

 Recall that the (standardized or not) Riesz $s$-energy  for a pair of particles a distance $r$ apart (see
\Ref{RieszVs}) takes its unique maximum for $s<0$ at $r=0$.
 Since all three pairwise distances vanish in the completely degenerate configuration, 
no other arrangement of particles can have a higher energy. 

Theorem \ref{absMAX} is proved. \hfill $\square$
\medskip

 So we have arrived at the following characterization: for $s<-2$,
the function $U_s(\gamma)$ has a relative minimum at $\gamma=0$ and an absolute maximum at $\gamma=\pi$;
moreover, at $\gamma=\pi/3$ it has a relative minimum when $-4<s<-2$ and a relative maximum when $s<-4$. 
 Moreover, earlier we already showed that the non-universal $\gamma_*(s)\in (0,\pi/3)$ iff $-4<s<-2$, so
it follows that for $-4<s<-2$ the non-universal equilibrium with $\gamma_*(s)\in (0,\pi/3)$ is a 
relative maximizer, with $U_s(\gamma_*(s)) > U_s(0)$, i.e. 
$\langle V_s\rangle \big(\tfrac{\pi-\gamma_*(s)}{2},\tfrac{\pi-\gamma_*(s)}{2},\gamma_*(s)\big) > 
\langle V_s\rangle \big(\tfrac{\pi}{2},\tfrac{\pi}{2},0\big) $
 if $-4<s<-2$.
 This proves Lemma 1 for $-4<s<-2$.

 Next, at $s=-4$ there is no non-universal equilibrium, so 
the function $U_{-4}(\gamma)$ has a relative minimum at $\gamma=0$ and an absolute maximum at $\gamma=\pi$;
moreover, at $\gamma=\pi/3$ it has a saddle point. 
 Therefore, at $s=-4$ the antipodal arrangement is the only, hence the absolute, minimizer. 
 Furthermore, the function $\gamma\mapsto U_{-4}(\gamma)$ is monotonically increasing for $\gamma\in[0,\pi]$.

 Next, it follows right away from \Ref{UsOFgamma} that for any fixed $\gamma\in[0,\pi]$, the map
$s\mapsto U_s(\gamma)$ is strictly monotonically decreasing on $s\in(-\infty, -2)$. 
 Therefore, since earlier we already showed that the non-universal $\gamma_*(s)\in (\pi/3,\pi/2)$ iff $s<-4$, 
it follows that for $-\infty<s<-4$ the non-universal equilibrium with $\gamma_*(s)\in (\pi/3,\pi/2)$ has 
$U_s(\gamma_*(s)) > U_{-4}(\pi/3)>U_s(0)=-2$ for $s<-4$, and therefore
$\langle V_s\rangle \big(\tfrac{\pi-\gamma_*(s)}{2},\tfrac{\pi-\gamma_*(s)}{2},\gamma_*(s)\big) > 
\langle V_s\rangle \big(\tfrac{\pi}{2},\tfrac{\pi}{2},0\big)$ if $s<-4$.
 This proves Lemma 1 for $s<-4$.

Lemma \ref{nonUNIVenergyBOUND} is proved. \hfill $\square$

 We now have all ingredients to finish the proof of Theorem \ref{absMIN}.
 Namely, by Lemma \ref{nonUNIVenergyBOUND}, for $s<-2$ the average pair energy of the non-universal configuration
$\langle V_s\rangle \big(\tfrac{\pi-\gamma_*(s)}{2},\tfrac{\pi-\gamma_*(s)}{2},\gamma_*(s)\big)$
is always strictly larger than the smaller of 
$\langle V_s\rangle \big(\tfrac{\pi}{3},\tfrac{\pi}{3},\tfrac\pi3\big)$
and
$\langle V_s\rangle \big(\tfrac{\pi}{2},\tfrac{\pi}{2},0\big)$.
 Therefore, and since the completely degenerate configuration yields the absolute maximizer, 
to determine the absolute minimizer for $s<-2$ only the latter two energy functions need to be compared.
 By an elementary calculation one now finds that the equilateral configuration is the absolute minimizer
for $s> s_3 = \frac{\ln(4/9)}{\ln(4/3)}$, while the antipodal arrangement is the absolute minimizer
for $s< s_3$.

Theorem \ref{absMIN} is proved. \hfill $\square$
\begin{remark}
 The fact that the equilateral configuration is the absolute $N=3$ minimizer for all $s>-2$ is a special 
case of a more general theorem on universal minimizers by Cohn and Kumar \cite{CoKu2007}.
\end{remark}

 We will next prove Theorem \ref{rel}, then Theorem~\ref{sad}.
\begin{remark}
 For the proofs of Theorems \ref{rel} and \ref{sad}, Fig.\ref{UsGRAPHS} is of only limited help. 
 True, Fig.\ref{UsGRAPHS} correctly shows that the equilateral triangle, at $\gamma_*=\pi/3$, is the only 
relative minimizer for $s>\frac{\ln(4/9)}{\ln(4/3)}=s_3$, and then also absolutely minimizing. 
 It also shows that at $\gamma_*=\pi/3$ is a saddle point for $s=-4$. 
 And Fig.\ref{UsGRAPHS} correctly shows that the antipodal arrangement is the absolute minimizer for $s < s_3$, and that the absolute
maximum occurs at $\gamma_*=\pi$, independently of $s$.
 Yet any other question of relative maxima or minima which are not absolute, or those of the saddle points,
cannot be answered by inspecting Fig.\ref{UsGRAPHS}.
 For this we need to study
 $\langle V_s\rangle (\alpha,\beta,\pi-\alpha-\beta)$ 
with 
$(\alpha,\beta)\in[0,\pi]^2\cap \{\alpha+\beta \leq \pi\}$ to
reveal the variations of the energy also for non-isosceles triangles. 
\end{remark}

\noindent
\textit{Proof of Theorem \ref{rel}:} 

 Turning next to the antipodal arrangement for $s<-2$, we compute the
Hessian of $\langle V_s\rangle (\alpha,\beta,\pi-\alpha-\beta)$, denoted $H_s(\alpha,\beta)$, 
evaluated at $\alpha=\frac{\pi}{2} =\beta$, and find essentially the identity matrix, viz.
\begin{equation}\label{HessianANTIpodal}
H_s\big(\tfrac\pi2,\tfrac\pi2 \big)= \frac{1}{3\cdot 2^s}\left(\begin{array}{lr} 1 &0\cr  0  &1\end{array}\right);\quad s<-2.
\end{equation}
 Therefore  the antipodal arrangement is a local energy minimizer for $s<-2$.

 Lastly, we compute $H_s(\alpha,\beta)$, 
evaluated at $\alpha=\frac{\pi}{3} =\beta$, and find 
\begin{equation}\label{HessianEQUILATERAL}
H_s\big(\tfrac\pi3,\tfrac\pi3 \big)= 
\frac{4+s}{9\surd3^s}\left(\begin{array}{lr} 2 & 1\cr 1  &2\end{array}\right);\quad s\in\Rset.
\end{equation}
 Since the matrix $\left(\begin{array}{lr} 2 & 1\cr 1  &2\end{array}\right)$ has  eigenvalues 1 and 3, it is positive definite, 
and so it follows that $H_s\big(\tfrac\pi3,\tfrac\pi3 \big)$ is positive definite for $s>-4$ and negative
definite for $s<-4$. 
 This proves that the equilateral configuration is a local energy minimizer for $s>-4$, and a local energy maximizer
for $s<-4$, as claimed.

Theorem \ref{rel} is proved. \hfill $\square$
\newpage

\noindent
\textit{Proof of Theorem \ref{sad}:} 

 We distinguish the regime $s<-1$ where every force equilibrium is proper, and the regime $s\geq -1$ with
its two pseudo force equilibria.

 Beginning with the regime $s\geq -2$, we know that the only equilibria are the universal arrangements. 
 We also know that the equilateral configuration is the absolute minimizer for all $s\geq -2$, and that the 
completely degenerate configuration is the absolute maximizer for $-2\leq s <0$. 
 So we only need to determine the character of the antipodal arrangement for $-2\leq s <0$.
 But this pseudo equilibrium is easily seen to be a saddle point: (i) keeping the two coincident particles 
fixed and moving the remaining one from its antipodal position monotonically to the position of the coincidental
particles will manifestly  increase the Riesz $s$-energy, and monotonically so; (ii) keeping the noncoincidental
particle fixed but now separating the two other particles from their coincidental position in an isosceles manner,
and monotonically so, will at first decrease the Riesz $s$-energy until one reaches the equilateral configuration.
 For suppose not, then this isosceles deformation of the antipodal arrangement would have to increase the energy before 
meeting the equilateral configuration at its absolute minimum energy --- and then there had to be a relative maximum 
in between, which is impossible. 
 Hence the antipodal $N=3$ configurations are saddle points for $-2<s<0$.

 Turning to the regime $s<-2$, all equilibria are proper and we can compute the Hessian of the energy function.

 We already proved that the equilateral configuration is absolutely minimizing for $s>s_3$, relatively
minimizing for $-4<s<s_3$, a saddle for $s=-4$, and relatively maximizing for $s<-4$.

 We also proved that the completely degenerate configuration is the absolute maximizer for $s <-2$,
and that the antipodal arrangement is absolutely minimizing for $s<s_3$ and relatively minimizing
for $s\in(s_3,-2)$. 

 It therefore remains to determine the character of the two non-universal families of isosceles equilibria.
 We have no closed formula for its Hessian, but the following argument seems compelling:
 Since the function $\langle V_s\rangle (\alpha,\beta,\pi-\alpha-\beta)$ is differentiable, the
min/max properties of the universal equilibria for $s<-2$ force the non-universal equilibria to be saddles;
for assume not, then by a mountain pass lemma it should follow that there has to be yet another equilibrium configuration ---
in contradiction to the fact that we have found all equilibria.

 Here is an elementary argument.

 First, if $s\in(s_3,-2)$ 
then the equilateral configuration and the antipodal arrangement are both local minimizers,
and the non-universal equilibrium is located on the segment of the height between the center and a midpoint of a side of 
the fundamental triangle, with the energy viewed as a function along the height having a local maximum at the non-universal
triangle; cf. Fig.\ref{UsGRAPHS}.
 Now suppose this non-universal equilibrium were a local energy maximizer also against arbitrary variations of the triangular shape. 
 Then keeping $\gamma_*(s)$ fixed and setting $\beta = \pi - \gamma_*(s)-\alpha$ the average Riesz $s$-pair energy function 
$\alpha\mapsto\langle V_s\rangle (\alpha,\pi-\gamma_*(s)-\alpha,\gamma_*(s))$ with  $\alpha\in(0,\pi-\gamma_*(s))$
must have a maximum at $\alpha = \frac{\pi-\gamma_*(s)}{2}$. 
 This is equivalent to the function 
$\alpha\mapsto W_s(\alpha) \equiv -\sin^{|s|}(\alpha) - \sin^{|s|}(\gamma_*(s)+\alpha)$ with  $\alpha\in(0,\pi-\gamma_*(s))$
having a maximum at $\alpha = \frac{\pi-\gamma_*(s)}{2}$. 
 However, by a discussion very similar to the one in the proof of Proposition \ref{equilISO} one can easily show that $W_s(\alpha)$ 
has a minimum at $\alpha = \frac{\pi-\gamma_*(s)}{2}$ as long as $s\in(s_3,-2)$ --- contradiction. 

 Similarly, if $s\in(-\infty,s_3)$ then the equilateral configuration and the completely degenerate one-point arrangement are both 
local maximizers, and the non-universal equilibrium is located on the segment of the height between the center and a corner of
the fundamental triangle, with the energy viewed as a function along the height having a local minimum at the non-universal
triangle; cf. Fig.\ref{UsGRAPHS}.
 Now suppose this non-universal equilibrium were a local energy minimizer also against arbitrary variations of the triangular shape. 
 Again keeping $\gamma_*(s)$ fixed and setting $\beta = \pi - \gamma_*(s)-\alpha$ we conclude that the function 
$\alpha\mapsto W_s(\alpha)$ defined above, with  $\alpha\in(0,\pi-\gamma_*(s))$, must have a minimum at $\alpha = \frac{\pi-\gamma_*(s)}{2}$. 
 Yet this time a discussion very similar to the one in the proof of Proposition \ref{equilISO} readily reveals that $W_s(\alpha)$ 
has a maximum at $\alpha = \frac{\pi-\gamma_*(s)}{2}$ as long as $s\in(\infty,s_3)$ --- contradiction. 

Theorem \ref{sad} is proved. \hfill $\square$

\begin{remark}\label{compactification}
Recall that for $s\geq 0$ the energy of any degenerate configuration is $+\infty$ and so we cannot meaningfully 
speak of relative or absolute maxima of the energy function. 
 If we compactly compress the pair energy as described in the introduction, then the completely degenerate
configuration is the absolute maximizer also for $s\geq 0$, while the antipodal arrangement is a saddle. 
\end{remark}\vspace{-20pt}
%
	\section{Energy landscapes: A monkey saddle}\vspace{-10pt}

 When graphing the average pair energy $\langle V_s\rangle (\alpha,\beta,\gamma)$ 
over the equilateral triangular domain $(\alpha,\beta,\gamma)\in[0,\pi]^3\cap \{\alpha+\beta + \gamma = \pi\}$ 
(which we have called the ``\emph{fundamental triangle}'' in $(\alpha,\beta,\gamma)$ space), for each
value of the parameter $s$ the graph shows a landscape with a threefold symmetry with
the equilateral equilibrium configuration $(\alpha,\beta,\gamma)=(\frac\pi3,\frac\pi3,\frac\pi3)$ at its center.\footnote{Of 
  course, to literally plot this graph in a cartesian diagram we would need four space dimensions, which
  we cannot ``see'' simultaneously. 
  However, one could visualize this graph as a two-dimensional surface over a two-dimensional planar equilateral 
  triangular domain embedded in three dimensions by using the $(1,1,1)$ direction as the axis for $\langle V_s\rangle (\alpha,\beta,\gamma)$.}
 This symmetry is simply due to the permutation invariance of $\langle V_s\rangle (\alpha,\beta,\gamma)$ 
w.r.t. the three angles.
 In particular, when $s\in(-\infty,-4)\cup(-4,-2)$, 
there are three absolute maxima [at $(\pi,0,0)$, $(0,\pi,0)$, and $(0,0,\pi)$] and three
minima [at $(\frac\pi2,\frac\pi2,0)$, $(\frac\pi2,0,\frac\pi2)$, and $(0,\frac\pi2,\frac\pi2)$], plus 
a critical point at its center $(\frac\pi3,\frac\pi3,\frac\pi3)$ which is a relative maximum for 
$s\in(-\infty,-4)$ and a relative minimum for $s\in(-4,-2)$, in either case of which there are also three
regular saddle points  associated with the non-universal equilibrium configuration, located on the heights either
between center and a midpoint of a side, or between center and a corner.
 
\begin{figure}[H]
\centering
\includegraphics[scale=.5]{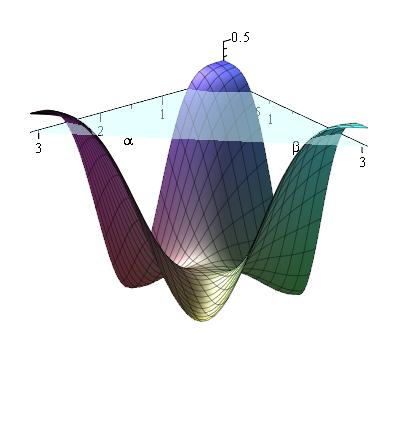}\vspace{-25pt}
\caption{Graph of $\langle V_{-4}\rangle (\alpha,\beta,\pi-\alpha-\beta)$ vs. $\alpha$ and $\beta$.}\vspace{-15pt}
\label{HilbertA}
\end{figure}

 When $s=-4$, which is the switching point for the equilateral equilibrium from being a relative maximizer to being
a relative minimizer (as $s$ increases through $-4$), the three regular saddle points merge at the center and 
degenerate into a single higher-order saddle point: a ``monkey saddle.''
 In Fig.\ref{HilbertA} we show the graph of $\langle V_{-4}\rangle (\alpha,\beta,\pi-\alpha-\beta)$ 
over the projection of the fundamental triangle into the $(\alpha,\beta)$ plane:
the isosceles triangular domain $(\alpha,\beta)\in[0,\pi]^2 \cap \{\alpha+\beta \leq \pi\}$;
note that this projected illustration somewhat distorts the three-fold symmetry.

 To the best of our knowledge, Hilbert and Cohn-Vossen \cite{HilbertCohnVossen} popularized 
the name ``monkey saddle'' for this type of higher order saddle, which is sometimes referred to as
``Hilbert's monkey saddle'' (see Fig.21 in \cite{Andersson}\footnote{Note that ``links'' and ``rechts'' are mixed up
  in the caption to Fig.21 of \cite{Andersson}.}).
 The perhaps most well-known monkey saddle with three-fold symmetry is the graph over the $x,y$ plane 
of the function $f(x,y) = \Re\esp(x+iy)^3$. 
 This illustrates a surface of (almost everywhere) negative Gauss curvature. 
 It is gratifying to see a monkey saddle surfacing naturally in the problem of 
the repulsive power-law force equilibrium of $N=3$ point particles on $\Sset^1$, which is 
embedded in the more interesting and important quest for $N$-particle equilibria on $\Sset^d$.

\begin{figure}[H]
\centering
\includegraphics[scale=.55]{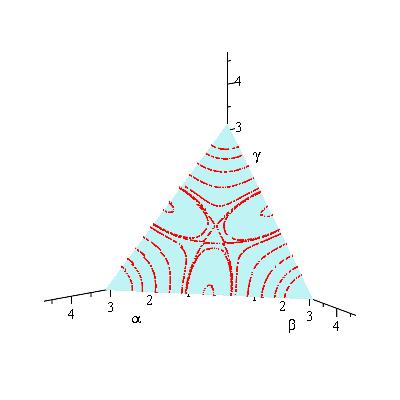} \vspace{-1.8truecm}
\caption{Contours of $\langle V_{-5}\rangle(\alpha,\beta,\gamma)$ in the fundamental $(\alpha,\beta,\gamma)$ triangle.}\vspace{-.9truecm}
\label{sMINfive}
\end{figure}

\begin{figure}[H]
\centering
\includegraphics[scale=.55]{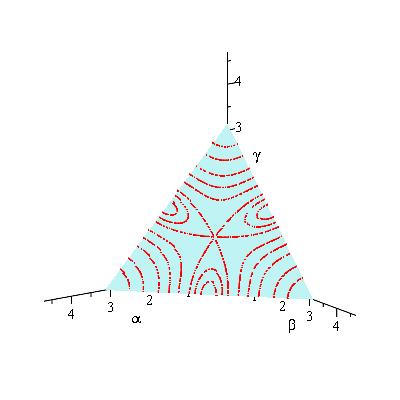} \vspace{-1.8truecm}
\caption{Contours of $\langle V_{-4}\rangle(\alpha,\beta,\gamma)$ in the fundamental $(\alpha,\beta,\gamma)$ triangle.}\vspace{-.9truecm}
\label{HilbertB}
\end{figure}

\begin{figure}[H]
\centering
\includegraphics[scale=.55]{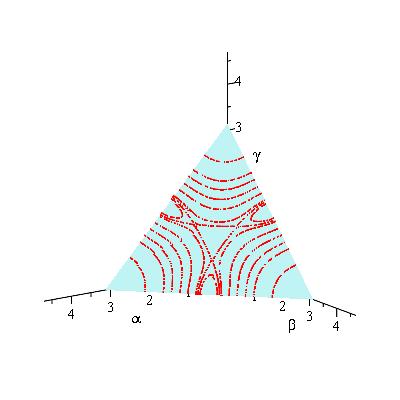} \vspace{-1.8truecm}
\caption{Contours of $\langle V_{-3}\rangle(\alpha,\beta,\gamma)$ in the fundamental $(\alpha,\beta,\gamma)$ triangle.}\vspace{-.9truecm}
\label{sMINthree}
\end{figure}

We close this section with three contour plots of $\langle V_{s}\rangle (\alpha,\beta,\gamma)$ in the
fundamental triangle of $(\alpha,\beta,\gamma)$ space for $s=-5$ (Fig.\ref{sMINfive}), $s=-4$ (Fig.\ref{HilbertB}),
and $s= -3$ ( Fig.\ref{sMINthree}). 


\newpage

\vspace{-20pt}

	\section{Summary and Outlook}\vspace{-10pt}
 
 Riesz $s$-energy minimization by degenerate configurations with all $N$ particles distributed over two antipodal points 
occurs for all\footnote{Of course, for $N=2$ the antipodal configuration is
trivially the absolute Riesz $s$-energy minimizer, but in this case, and only in this case, it is not a degenerate configuration.} $N>2$ when $s$ is ``negatively large enough'' (certainly $s<-2$); see Theorem 7 in \cite{Bjorck}.
 In particular, as pointed out by Rachmanov, Saff, and Zhou \cite{RSZa}, the results of 
 \cite{Bjorck} imply for even $N$ that the arrangement with 
half of the particles each placed at two antipodal points is the Riesz $s$-energy minimizer for all $s\leq -2$,
uniquely so if $s<-2$.
 When $N$ is odd, then antipodal minimizing arrangements seem to be absolutely minimizing for
$s\leq s_N<-2$, but very little seems known about $s_N$.
 Even less is known about equilibria in general.

 In this work we have studied the problem for the smallest odd $N=3$ for which it is interesting (clearly, $N=1$ is not!).
 In particular, we have given a detailed rigorous proof of the statement in Appendix 1 of \cite{NBK},
that the equilateral configuration is the unique absolute minimizer iff
$s> \frac{\ln(4/9)}{\ln(4/3)}$, while the antipodal arrangement with one particle in the North and two in the
South Pole is the unique absolute minimizer iff $s<\frac{\ln(4/9)}{\ln(4/3)}$. 

 Beyond the identification of the absolute minimizers, we have rigorously identified all repulsive power-law force equilibria, 
proper or pseudo, for $N=3$ particles on $\Sset^1$, and therefore also on $\Sset^d$, $d\in\Nset$. 
 We have also analyzed their relative energies, and their stability vs. small perturbations on $\Sset^1$. 
 We recall that the $N=3$ equilibria, and the absolute minimizers and maximizers amongst them, are the same for all $\Sset^d$, but the
stability of the not absolutely minimizing or maximizing configurations generally depends on $d$.
 Beside the universal equilibria, which are independent of $s$ and ``obvious to everyone,'' we have proved 
the existence of two ``not-so-obvious''  $s$-dependent continuous families of non-universal isosceles triangular equilibria.
 We found that both these families bifurcate off of the equilateral configuration at $s=-4$ in an analytical manner;
one family in addition bifurcates off of the antipodal arrangement at $s=-2$ in a non-analytical manner. 

 Our bifurcation study has also revealed two curiosities:
one is that at $s=-4$, the point of the analytical bifurcation, 
the energy landscape degenerates into the shape of a monkey saddle;
the other is the peculiar non-analytical bifurcation at $s=-2$, in the neighborhood of which two solution branches 
converge toward each other faster than any inverse power when $s\uparrow -2$.
 This type of function is usually used in calculus courses as an example of a $C^\infty$ function which has a 
Taylor series about the non-analytical point with an infinite radius of convergence, yet it converges to the 
defining non-analytical function only at the non-analytical point. 
 Inevitably this type of example function has always had the ``smell of being cooked up'' for the purpose.
 As with the monkey saddle, it is gratifying to see such a non-analytical function show up
naturally in the context of an interesting problem.

 Similarly complete studies might be possible also for $N=5$, 
and perhaps $N=7$, but the complexities will soon become overwhelming.

 For instance, even the absolutely energy-minimizing $5$-particle arrangement on $\Sset^d$ 
changes several times when $s$ varies over the real line, apparently:

\noindent
($\star$) For $s<-2$, all minimizers are either triangular or antipodal \cite{Bjorck}, independently of $d$.
 In Appendix 1 of \cite{NBK} it is reported that their own computer-assisted results showed that
an antipodal arrangement of two point particles at the South and three at the North Pole is the optimizer for $s \leq -2.368335...$; 
at $s=-2.368335...$ a crossover takes place, and for $-2.368335... \leq s \leq  -2$  an isosceles triangle on a great circle, with 
one particle in the North Pole and two particles each in the other two corners, with (numerically) optimized height, is 
an energy-minimizing arrangement of five point particles.
 This has yet to be proved rigorously. 
 Also, the question of general equilibria for $s<-2$ has not yet been answered.

\noindent
($\star$) For $s = -2$ a whole family of configurations satisfying $\sum_{i=1}^5 \pV_i = \mathbf{0}$, 
can be shown with elementary arguments to be minimizers, but the possibilities depend on $d$.
	For instance, when $d=2$, 
at $s=-2$ the isosceles $\Sset^1$ minimizer described in the previous paragraph is the degenerate 
limit of a continuous family of rectangular pyramids with height $h=5/4$ which contains the square pyramid as special 
case, all of which have the same energy $-3/4$ at $s=-2$.
 At $s=-2$ the square pyramid can also be continuously deformed on $\Sset^2$ into
a regular triangular bi-pyramid without changing the Riesz $-2$-energy.

\noindent
($\star$) For $s > -2$ the minimizers depend on $d$, too.
 Thus, the regular pentagon is the absolute minimizer of $N=5$ particles on $\Sset^1$ when $s>-2$ (cf.
\cite{Fekete}), while the $5$-point simplex is (presumably) the $s>-2$ optimizer on $\Sset^d$ whenever $d\geq 3$.
 The case $d=2$ is more complicated, however!
 Summarizing numerical results for a closely spaced selection of values $s\in[1,400]$ reported in \cite{MeKnSm1977}, 
and of \cite{BermanHanes} for $s=-1$, in Appendix 1 of \cite{NBK} it is reported that 
(in the notation of \cite{Sch2016arXiv}) for $-2 < s < \shin$, with $\shin = 15.048077392\dots$,
a regular triangular bi-pyramid seems to be the unique (up to rotation) energy-minimizing configuration.
	At $s = \shin$  a crossover happens, at which the triangular bi-pyramid and 
a square pyramid with height $h\approx 1.1385$ have the same Riesz $s$-energy, and
a square pyramid with (numerically) optimized $s$-dependent height\footnote{For $-2\leq s\leq 0$ the optimal 
          height of the square-pyramidal configuration is constant, equal to $5/4$. 
          The optimized height depends on $s$ only for $s>0$.\vspace{-20pt}}
seems to be the energy-minimizing configuration for real $s > \shin $ (at least up to $s=400$, cf. \cite{MeKnSm1977}).
        So far, using elementary arguments, the regular triangular bi-pyramidal configuration was rigorously shown 
in \cite{DrLeTo2002} to be the unique (up to rotation) optimizer on $\Sset^2$ of the logarithmic energy ($s=0$).
        A rigorous, computer-assisted proof of the optimality on $\Sset^2$ of the regular triangular bi-pyramidal configuration 
if $s\in (-2,0)$ or $s\in (0, \shin)$ was achieved only recently \cite{Sch2016arXiv}, where it is also proved that the 
regular triangular bi-pyramidal configuration is not an optimizer on $\Sset^2$ when $s>\shin$.
        That the same configuration maximizes the sum of distances on $\Sset^2$ and therefore minimizes the
Riesz $-1$-energy on $\Sset^2$ was established earlier with a different computer-aided proof in \cite{HouSh2011}. 
        Reference \cite{Sch2016arXiv} also contains a computer-assisted proof that a square pyramid 
with $s$-dependent height is the unique (up to rotation) optimizer on $\Sset^2$ when $\shin < s < 15+25/512$.

\noindent
($\star$) At ``$s=\infty$'' another ``crossover,''  or rather a ``continuous degeneracy,'' happens.
 The triangular bi-pyramid and the square pyramid with height $1$ both are particular best-packing configurations,
which is rigorously known.

        Many of these computer-assisted results still await their rigorous proof without the assistance of
a machine.
        The hard part is to show that one hasn't overlooked any relative minimizer which 
competes with the known ones.
 \smallskip

\noindent
\textbf{Acknowledgment}:\!\! We thank\! Johann Brauchart for\! \cite{Andersson}\! and his comments,\hspace{-5pt}
and Bernd Kawohl for the history of the book by Hilbert \&\ Cohn-Vossen.
 We also thank the referee for the constructive criticism. 

\bibliographystyle{modamsplain}

\end{document}